\documentclass[aps,prc,preprintnumbers,12pt,showpacs]{revtex4-1}
\usepackage{natbib}
\usepackage{amsmath}
\usepackage{graphicx}
\usepackage{hyperref}
\usepackage{slashbox}
\usepackage{longtable}
\usepackage{color}
\usepackage{bm}

\begin{document}

\preprint{JLAB-THY-19-3012}

\title{Nuclear Theory and Event Generators for Charge-Changing Neutrino Reactions}
	
\author{J.~W.~Van Orden} 
\affiliation{Department of Physics, Old Dominion University, Norfolk, VA 23529\\ Jefferson Lab,12000 Jefferson Avenue, Newport News, VA 23606, USA 
\footnote{Notice: Authored by Jefferson Science Associates, LLC under U.S. DOE Contract No. DE-AC05-06OR23177.
The U.S. Government retains a non-exclusive, paid-up, irrevocable, world-wide license to publish or reproduce this manuscript for U.S. Government purposes}
}
\author{T.~W.~Donnelly}
\affiliation{Center for Theoretical Physics, Laboratory for Nuclear Science and Department of Physics, Massachusetts Institute of Technology, Cambridge, MA 02139, USA}

\date{\today}

\begin{abstract}

Semi-inclusive $CC\nu$ cross sections based on factorized cross sections are studied for a selection of spectral function models with the objective of facilitating the choice of models for use as input into event generators. The basic formalism for such cross sections is presented along with an introduction to constructing spectral functions for simple models based on the independent particle shell model (IPSM), the relativistic Fermi gas model (RFG) and a local density approximation (LDA) based on the RFG. Spectral functions for these models are shown for $^{16}$O along with a more sophisticated model which includes nucleon-nucleon interactions \cite{ALVAREZRUSO20181}. Inclusive and semi-inclusive cross sections are calculated for these models. Although the inclusive cross sections are all of similar size and shape, the semi-inclusive cross sections are subtantially different depending upon whether the spectral functions contain some features associated with the nuclear shell model or are based on the RFG and LDA models. Calculations of average values and standard deviations of the initial neutrino energy using the semi-inclusive cross section for the various models are presented and indicate that there may be simple kinematical descriptions of the average neutrino energy which is common to all of these models.
\end{abstract}


\maketitle
	
\section{Introduction}\label{sec:intro}

Accelerator-based neutrino scattering experiments provide an important tool for determining parameters of the standard model and for exploring possible physics beyond the standard model \cite{ALVAREZRUSO20181}. Due to the small size of weak interaction cross sections, obtaining reasonable counting rates for these experiments requires large amounts of target material. As a result, most experiments rely on readily available materials, namely those containing water or hydrocarbons and hence hydrogen, carbon or oxygen nuclei. Additionally, some heavier nuclei, such as argon and iron,  are sometimes employed. As a result analysis of such experiments at high energies requires some knowledge of nuclear structure and nuclear reactions in order to extract the required information on neutrino reactions with individual nucleons and to determine the incident neutrino energy of measured events, since the neutrino production mechanism typically results in beams with a very broad flux distribution, having widths typically measured in GeV.

Given the relatively low counting rates, it is typical for experiments to bin events to produce either {\em inclusive} cross sections, namely those initiated by an incident neutrino with only the final-state charged lepton detected, or cross sections containing all such events except those with detected pions, denoted CC$\nu$ and CC$\nu$0$\pi$ reactions, respectively. Nuclear theorists who are working to assist in the interpretation of neutrino reactions in the GeV regime have therefore tended to concentrate on the calculation of this class of cross sections. The models employed may contain contributions from quasielastic (QE) scattering where the neutrino reaction is assumed to result in the ejection of a single nucleon, contributions to the scattering due to two-body currents and short-range correlations which can produce two nucleons in the final state, or in events where mesons are produced directly through background processes, nucleon resonances and by deep inelastic scattering \cite{ALVAREZRUSO20181}. Work has also been done on studying the effect of production of collective nuclear resonance states as calculated via the random phase approximation (RPA). Furthermore, as an indication of possible improvements in treating the nuclear many-body problem, {\it ab initio} calculations --- albeit using non-relativistic dynamics --- of QE CC$\nu$ scattering from carbon have been performed by means of large-scale quantum Monte Carlo methods. Such inclusive reactions involve total hadronic cross sections and typically are relatively insensitive to the details of the final nuclear states reached; accordingly rather simple models may yield cross sections that are not very different from those found in the most sophisticated models. Typically, as long as the essential aspects of relativistic kinematics and incorporation at a reasonable level of unitarity (and hence the sum rules this entails) are taken into account, the inclusive predictions using dramatically different models are rather similar and  which agree to about 10-20\% \cite{ALVAREZRUSO20181}.

A phenomenological scaling approach to predicting the inclusive CC$\nu$ cross sections has also been pursued. This relies on the fact that semi-leptonic electroweak processes are closely related and hence that a set of scaling functions can be determined from analyses of inclusive electron scattering measurements, which can then be used to incorporate the basic needed nuclear response in studies of CC$\nu$ reactions at comparable energies. The most elaborated version of such scaling approaches is the so-called SuSAv2+MEC model \cite{Megias:2017cuh} which has been shown to represent the electroweak inclusive cross sections over a wide range of energies, excluding very near threshold where the approach is not designed to be applicable. 

\begin{figure}
	\centerline{\includegraphics[height=2.5in]{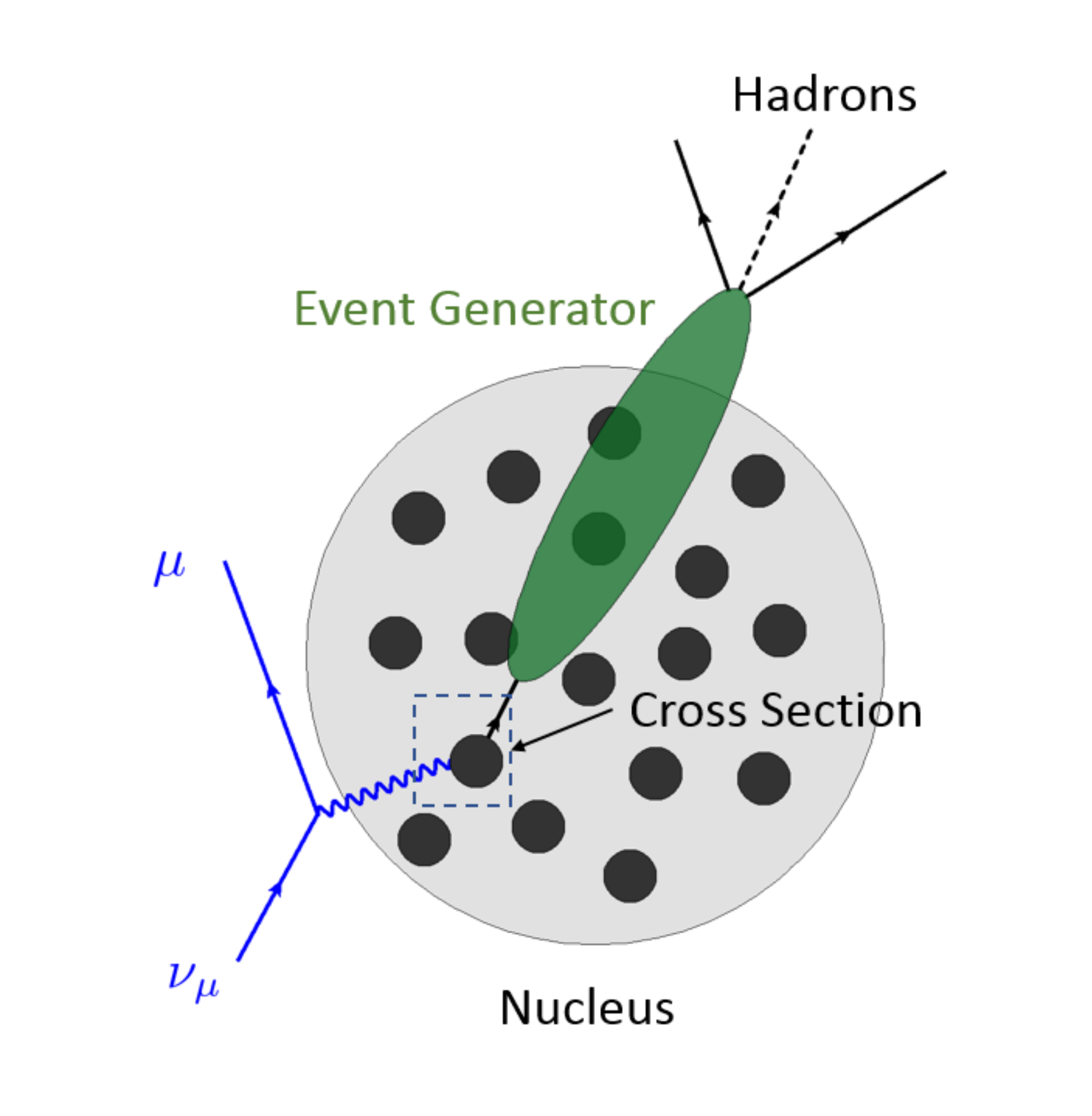}}
	\caption{(Color online) A schematic representation of a typical event generator. The scattering event contained within the dashed box represents the initial semi-inclusive cross section. The ellipse represents the method used by the event generator to propagate the hadrons produced by the initial nuclear event through the residual nucleus. }\label{fig:event_generator}
\end{figure}

In contrast to the theoretical issues summarized above for inclusive (total hadronic) cross sections, modern experimental studies of CC$\nu$ reactions rely on the use of data simulations to determine the behavior of the detectors involved and to provide predictions for the measured data. The initial part of the simulation process involves an event generator to provide predictions of events covering the available phase space that can be used as input to programs that model the characteristics of the specific detectors. A schematic representation of a typical event generator is shown in Fig.~\ref{fig:event_generator}. The event generator proceeds in the following manner. One starts by computing the cross section for CC$\nu$ scattering from a nucleon in the target nucleus using some model for the primary electroweak process. To date this typically means employing some simple model that has been used to describe the inclusive reaction. The nucleon and other possible hadrons produced by this event are then propagated through the rest of the nucleus by means of a cascade model or other statistical approximation to produce the final distribution of these hadrons. This rather crude approach is necessitated by the fact that reliable, consistent calculations of all of the possible final state channels are beyond the current capabilities of nuclear theory. Even if such calculations where possible, the computer time necessary to perform them would preclude the production of the large number of events needed for an effective simulation of data.  

Given the crudeness of this approach being employed in typical event generators one should not expect more than rather integrated quantities such as the inclusive cross section to be reliably simulated. In recent studies, however, it has become common to employ events in which not only the final-state charged lepton is detected, but also some hadron as well. This is motivated by the above rationale together with the desire to constrain the kinematics and thereby to constrain the incident neutrino energy better than can be accomplished by detecting only the final-state charged lepton. However, even in a next-most-complicated situation where a proton is detected in coincidence with a final-state electron or muon, the primary reaction indicated in Fig.~\ref{fig:event_generator} is not an inclusive one, but is now a {\em semi-inclusive} reaction \cite{semi}. This is the analog of going from inclusive $(e,e')$ reactions to semi-inclusive $(e,e'p)$ reactions. For example, studies involving muon detection together with liquid argon TPCs to detect ejected protons yield this specific event class. 

Unfortunately, such a desire on the experimental side also necessitates that on the theoretical side one now must confront the much more complicated semi-inclusive reaction. Modeling of semi-inclusive electron scattering has been undertaken for several decades and from that experience one knows how much more difficult the problem becomes when any aspect of the final nuclear states reached in the reaction is required. On the one hand, where the energies typically involved in large-scale neutrino oscillation studies much lower, then the relatively small number of exclusive final states reached might prove to be tractable in future modeling. Indeed, the low-energy beam stop neutrino facilities do have this advantage. On the other hand, were the relevant energies much higher, then the typical high-energy physics approach of factorizing the problem into ``soft'' physics convoluted with ``hard'' perturbative physics might be motivated. Unfortunately, this is not the case: the typical energy regime used for practical neutrino oscillation studies is a few GeV where the problem is not simple for either reason.  

Accordingly, the primary process in Fig.~\ref{fig:event_generator} should typically be a CC$\nu$ reaction on a neutron in the target nucleus yielding a muon and proton in the final state, (perhaps) followed by propagation of the proton through the nucleus until it emerges and is detected (for instance, in an argon TPC). This primary process is not an inclusive one, but is semi-inclusive, and involves not the total hadronic cross section, but the specific asymptotic state that defines the event. Said another way, one should not expect the primary reaction to produce an intermediate state that then arranges itself into the event class of interest solely via propagation in the nucleus: indeed, the semi-inclusive reaction typically requires the quantum mechanical overlap of the many-body ground state of the nucleus and a non-trivial final nuclear state of the required current operators. This cross section depends on the measurement of five independent quantities in the final state, the magnitude of the momenta of nucleon $p_N$ and the muon $k'$, the angle of the muon three-momentum relative to the direction of the neutrino beam, the polar angle of the nucleon three momentum relative to the beam direction $\theta_N^L$ and the azimuthal angle of the nucleon three-momentum relative to the plane containing the beam direction and the muon three-momentum.   

Thus it is clear that structure of the event generator requires the calculation of semi-inclusive cross sections rather than the inclusive cross sections that have been the focus of most theoretical activity. In general, these cross sections may contain not only a single nucleon but also additional hadrons that can be produced directly by the neutrino scattering, something that is presently beyond the scope of high-energy nuclear theory.  Accordingly, the separation of the production of multiple hadrons in the primary interaction and those associated with the final-state interactions provided by the event generator is a topic the requires some discussion between nuclear theorists and the developers of the event generators. 

The objective of this paper is to study semi-inclusive CC$\nu$ cross sections produced by several of the nuclear models that have been used to produce single nucleons as input for quasielastic scattering in various event generators.
We will show that these models yield very similar descriptions of inclusive QE scattering and yet result in significantly different results for semi-inclusive scattering, and therefore that much more care must be taken in determining which models can be used reliably in this role. Specifically, in Sect.~\ref{sec:cross} we summarize the basic formalism required in treatments of semi-inclusive neutrino reactions; more details on these developments can be found in \cite{semi} and \cite{Megias:2017cuh}, as well as in two appendices in this paper. This is followed in Sect.~\ref{sec:models} by the introduction of four specific popular models to represent the needed spectral function that is defined in Sect.~\ref{sec:cross}, namely, in Sect.~\ref{subsec:IPSM} of the independent particle shell model, in Sect.~\ref{subsec:RFG} of the relativistic Fermi gas model, in Sect.~\ref{subsec:LDA} of the extension of the relativistic Fermi gas to the local density approximation, and in Sect.~\ref{subsec:Rome} of results obtained using a state-of-the-art spectral function. In Sect.~\ref{sec:results} we then employ these modeled spectral functions to obtain momentum density distributions (Sect.~\ref{subsec:momdist}), inclusive CC$\nu$ cross sections (Sect.~\ref{subsec:incl}), and semi-inclusive results (Sect.~\ref{subsec:semi}). In Sect.~\ref{sec:Neutrino_Energy} we examine the possible extraction of the incident neutrino energy from semi-inclusive cross sections. Finally, in Sect.~\ref{sec:concl} we state our conclusions and summarize what has been learned from this study. 

\section{CC$\nu$ Cross Sections Using Spectral Functions}\label{sec:cross}

From the previous discussion, it is clear that the event generators require {\em at least} semi-inclusive scattering cross sections in a factorizable form that separates the primary reaction cross section in which the weak interaction on a single nucleon in the nucleus produces a nucleon with the required energy and momentum from the quantity that arises from modeling the nuclear many-body problem and that captures the probablity that the event occurs, namely, the so-called spectral function defined below. In the crudest version of such an approach one might first ignore the final-state interactions of the produced nucleon and thus invoke the so-called plane-wave impulse approximation (PWIA). We restrict our attention to this particular version in the present work. Of course, one can go beyond this and perhaps incorporate final-state interactions vis the so-called distorted-wave impulse approximation (DWIA) or perhaps employ the statistical ideas being used in current event generators. However, our goal is not to develop the ``best'' semi-inclusive model at present, but rather to explore the consequences of using various models for the nuclear physics involved to see how similar or different the results can be for inclusive versus semi-inclusive reactions even at the level of the PWIA. Accordingly, the natural representation of such a cross section relies on the calculation of the nuclear spectral function that describes the probability finding a nucleon in a nucleus with given momentum (called the missing momentum $p_m$) and with a given excitation energy of the residual nuclear system (called the missing energy $E_m$). The spectral function $S(p_m,E_m)$ is defined such that
\begin{equation}
S(p_m,E_m)\Lambda^+(\bm{p})=\sum_{s}\left< \psi_A(P_A)\right.
\left| P^\mu,s;\psi_{A-1}(P_{A-1})\right>
\left< P^\mu,s;\psi_{A-1}(P_{A-1})\right.
\left|\psi_A(P_A)\right>\,,
\end{equation}
where $\psi_A(P_A)$ and $\psi_{A-1}(P_{A-1})$ represent the wave function of the target nucleus A-body nucleus and of the residual (A-1)-body nucleus, respectively. The four-momentum $P^\mu$ is that of the nucleon that will absorb the W-boson. The positive-energy projection operator $\Lambda^+(\bm{p})$ is necessary to construct the relativistic single-nucleon CC$\nu$ cross section. The spectral function is normalized such that
\begin{equation}
n(p_m)=\int_0^\infty dE_m\, S(p_m,E_m)\,,\label{eq:spectral_momentum}
\end{equation}
where $n(p_m)$ is the nuclear momentum density distribution for which we use the normalization
\begin{equation}
\mathcal{N}=\frac{1}{(2\pi)^3}\int_0^\infty dp_m\,p_m^2n(p_m)\,.\label{eq:spectral_norm}
\end{equation}
Here, $\mathcal{N}$ is the number of nucleons that are active in the scattering. For the case of CC$\nu$ reactions (CC${\bar\nu}$) this is the number of neutrons (protons) in the nucleus; unless specified otherwise we shall assume the former in the rest of the discussions.

\begin{figure}
	\centerline{\includegraphics[height=2.0in]{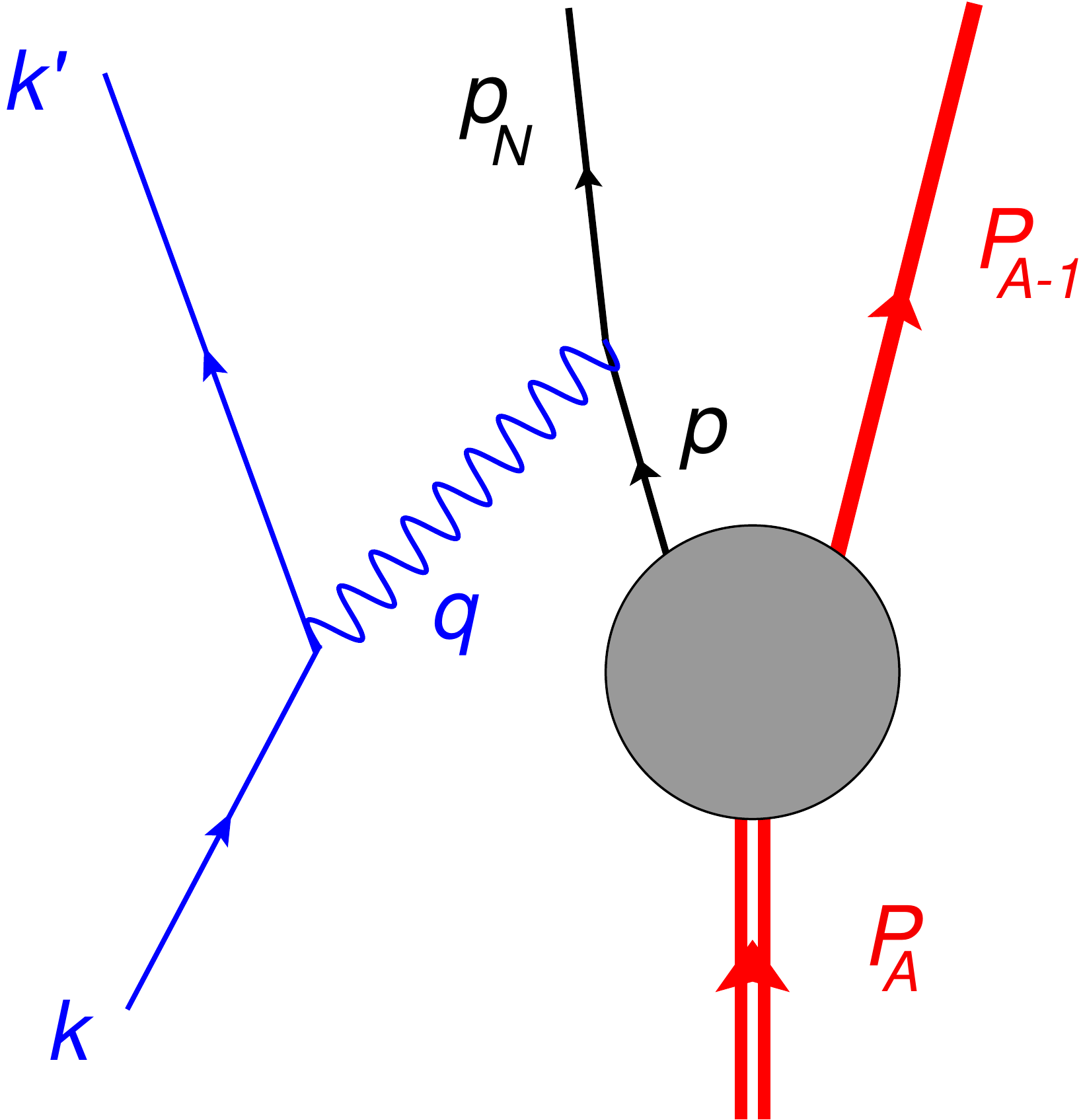}}
	\caption{(Color online) Feynman diagram representing the factorizable CC$\nu$ cross sections.}\label{fig:semi_inclusive}
\end{figure}

The amplitude for the semi-inclusive cross section in this separable approximation is represented by the Feynman diagram in Fig.~\ref{fig:semi_inclusive}.
Since the CC$\nu$ scattering will involve large momenta, it is necessary that the kinematics be treated relativistically. The initial neutrino four-momentum is given by
\begin{equation}
K^\mu=(\varepsilon,\bm{k})\,,
\end{equation}
where
\begin{equation}
\varepsilon=\sqrt{\bm{k}^2+m^2}
\end{equation}
and the final state muon four-momentum by
\begin{equation}
{K'}^\mu=(\varepsilon',\bm{k}')\,,
\end{equation}
where
\begin{equation}
\varepsilon'=\sqrt{{\bm{k}'}^2+{m'}^2}\,,
\end{equation}
with the neutrino and muon masses being $m$ and $m'$. The four-momentum of the target nucleus in its rest frame is
\begin{equation}
P_A^\mu=(M_A,\bm{0})\,,
\end{equation}
with ground-state rest mass $M_A$, the lab frame (L) four-momentum of the proton in the final state is
\begin{equation}
P_N^\mu=(\sqrt{\bm{p}_N^2+m_N^2},\bm{p}_N^L)
\end{equation}
and
the four-momentum of the residual nucleus is
\begin{equation}
P_{A-1}^\mu=(\sqrt{\bm{p}_m^2+W_{A-1}^2},\bm{p}_m)\,.
\end{equation}
Here $W_{A-1}$ is the invariant mass of the residual nucleus which is not in general in its ground state. The four-momentum of the struck nucleon is
\begin{equation}
P^\mu=P_A^\mu-P_{A-1}^\mu\,.
\end{equation}

\begin{figure}
	\centerline{\includegraphics[width=6.0in]{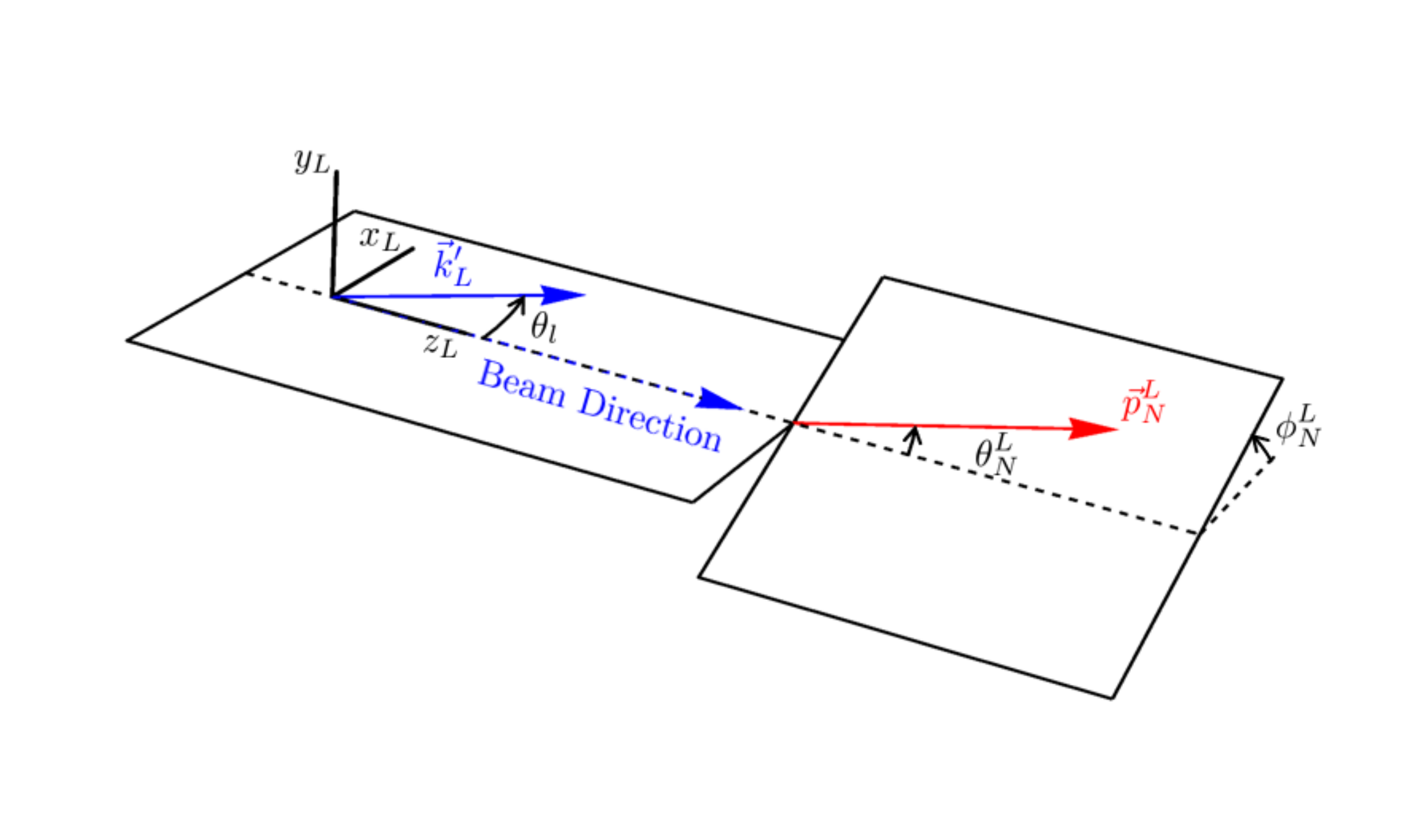}}
	\caption{(Color online) Kinematic variables in the lab frame where the beam direction is chosen as the z-axis.}\label{fig:coordinates}
\end{figure}

The cross sections are represented by kinematical variables defined relative the the lab frame coordinate system shown in Fig.~\ref{fig:coordinates}. The three-momenta defined in this frame are 	
\begin{align}
\bm{k}=&k\bm{e}_z\nonumber\\
\bm{k}'=&k'(\sin\theta_l\bm{e}_x+\cos\theta_l\bm{e}_z)\nonumber\\
\bm{p}_N^L=&p_N(\cos\phi_N^L\sin\theta_N^L\bm{e}_x+\sin\phi_N^L\sin\theta_N^L\bm{e}_y+\cos\theta_N^L\bm{e}_z)\,.
\end{align}

Using these definitions the semi-inclusive cross section in the factorization approximation is given by
\begin{align}
	\frac{d\sigma}{dk'd\Omega_{k'}dp_Nd\Omega_N^L}=&\frac{G_F^2\cos^2\theta_cm_N{k'}^2p_N^2}{8(2\pi)^6 \varepsilon'E_N}\int_0^\infty d\mathcal{E}\int_0^\infty dk\frac{P(k)}{k}\nonumber\\
	&\times\int d^3p_m v_0\widetilde{\mathcal{F}}^2_\chi S(p_m,E_m(\mathcal{E}))\delta\left(\bm{k}-\bm{k}'-\bm{p}_N^L-\bm{p}_m\right)\nonumber\\
	&\times \delta\left(M_A+\varepsilon-\varepsilon'-E_N-\sqrt{p_m^2+M_{A-1}^2}-\mathcal{E}\right)\,,\label{eq:semi_inclusive} 
\end{align}
where the integrations over $\mathcal{E}$ and $k$ are summations over the possible states of the residual nucleus and the possible values of the initial neutrino momentum that can contribute to the measured muon and nucleon momenta. The mass of the ground state of the residual nucleus is given by $M_{A-1}$. The variable $\mathcal{E}$ is difference in recoil energies for residual nucleus with invariant mass $W_{A-1}$ and that for a residual nucleus with minimum invariant mass $M_{A-1}$ and is defined as 
\begin{equation}
\mathcal{E}=\sqrt{\bm{p}_m^2+W_{A-1}^2}-\sqrt{p_m^2+M_{A-1}^2}\,.\label{eq:scriptE}
\end{equation}
$\widetilde{\mathcal{F}}^2_\chi$ is a reduced single nucleon cross section and $v_0$ is a kinematic factor. For completeness these are defined in Appendix \ref{sec:appA}. The function $P(k)$ is a flux factor that represents the weighting of the contributions of possible neutrino momenta associated with the momentum profile of the neutrino beam. The separation energy  $E_s=M_{A-1}+m_N-M_A$ is the minimun energy necessary to remove a nucleon from the nucleus. This implies that $M_A=M_{A-1}+m_N-Es$ and accordingly the energy-conserving delta-function can then be written as
\begin{equation}
D = \delta\left(\varepsilon-\varepsilon'-E_N+m_N-\sqrt{p_m^2+M_{A-1}^2}+M_{A-1}-\mathcal{E}-E_s\right)\,,
\end{equation}
namely, that the missing energy is the difference between the energy of the initial $A$-body nucleus plus the energy transfer from the lepton scattering and the energy of the detected proton. It can then be identified as
\begin{equation}
E_m=\mathcal{E}+E_s+\sqrt{p_m^2+M_{A-1}^2}-M_{A-1}\,. \label{eq:Em}
\end{equation} 
 Note that $\sqrt{p_m^2+M_{A-1}^2}-M_{A-1}$ is the recoil kinetic energy of the ground state of the $A-1$ system. Since $p_m$ is limited by the rapid fall of the spectral function in this variable, for all but the lightest nuclei $p_m \ll M_{A-1}$ and therefore, for detectable CC$\nu$ cross sections, $\sqrt{p_m^2+M_{A-1}^2}-M_{A-1}\cong 0$.  A further simplification  can be obtained by assuming that the incident neutrino is massless so that $\varepsilon=k$. With these approximations, the delta function in Eq.~(\ref{eq:semi_inclusive}) becomes
\begin{equation}
D \cong \delta\left(k-\varepsilon'-E_s-E_N+m_N-\mathcal{E}\right)\label{eq:energy_conservation}
\end{equation}
and $E_m\cong \mathcal{E}+E_s$\,.
The semi-inclusive cross section can then be written as
\begin{align}
\frac{d\sigma}{dk'd\Omega_{k'}dp_Nd\Omega_N^L}=&\frac{G_F^2\cos^2\theta_cm_N{k'}^2p_N^2}{8(2\pi)^6 \varepsilon'E_N}\int_0^\infty d\mathcal{E}\frac{P(k_0)}{k_0} v_0\widetilde{\mathcal{F}}^{2}_\chi S(p_m,E_s+\mathcal{E})\,.\label{eq:semi_simple}
\end{align}
The neutrino momentum is given by
\begin{equation}
k_0=\varepsilon'+E_s+E_N-m_N+\mathcal{E}\label{eq:k_zero}
\end{equation}
and the missing momentum is given by
\begin{align}
p_m=&\bigl[k_0^2+{k'}^2+p_N^2-2k_0k'\cos\theta_l-2k_0p_N\cos\theta_N^L\nonumber\\
& +2k'p_N(\cos\theta_l\cos\theta_N^L+\sin\theta_l\sin\theta_N^L\cos\phi_N^L)\bigr]^\frac{1}{2}\,.\label{eq:p_missing}
\end{align}

The corresponding inclusive cross section can be obtained by integrating Eq.~(\ref{eq:semi_inclusive}) over $\bm{p}_N^L$ to give
\begin{align}
\frac{d\sigma}{dk'd\Omega_{k'}}=&\frac{G_F^2\cos^2\theta_cm_N{k'}^2}{8(2\pi)^5 \varepsilon'}\int_0^\infty dk\frac{P(k)}{qk}\int_0^\infty d\mathcal{E} \int_0^\infty dp_mp_mv_0\widetilde{\mathcal{F}}^2_\chi S(p_m,E_s+\mathcal{E})\nonumber\\
&\times\theta(p_m^+-p_m)\theta(p_m-p_m^-)\,,\label{eq:inclusive}
\end{align}
where
\begin{equation}
p_m^+=\sqrt{(\omega-E_s-\mathcal{E})(\omega-E_s-\mathcal{E}+2m_N)}+q
\end{equation}
and	
\begin{equation}
p_m^-=\left|\sqrt{(\omega-E_s-\mathcal{E})(\omega-E_s-\mathcal{E}+2m_N)}-q\right|\,.
\end{equation}
The energy transfer is $\omega=k-\varepsilon'$.

\section{Simple Models of the Spectral Function}\label{sec:models}

We can now use Eqs.~(\ref{eq:semi_simple}) and (\ref{eq:inclusive}) to calculate cross sections resulting from the use of simple models by identifying the spectral functions for these models.

\subsection{Spectral Function for the Independent-Particle Shell Model (IPSM)}\label{subsec:IPSM}

\begin{figure}
	\centerline{\includegraphics[height=2.0in]{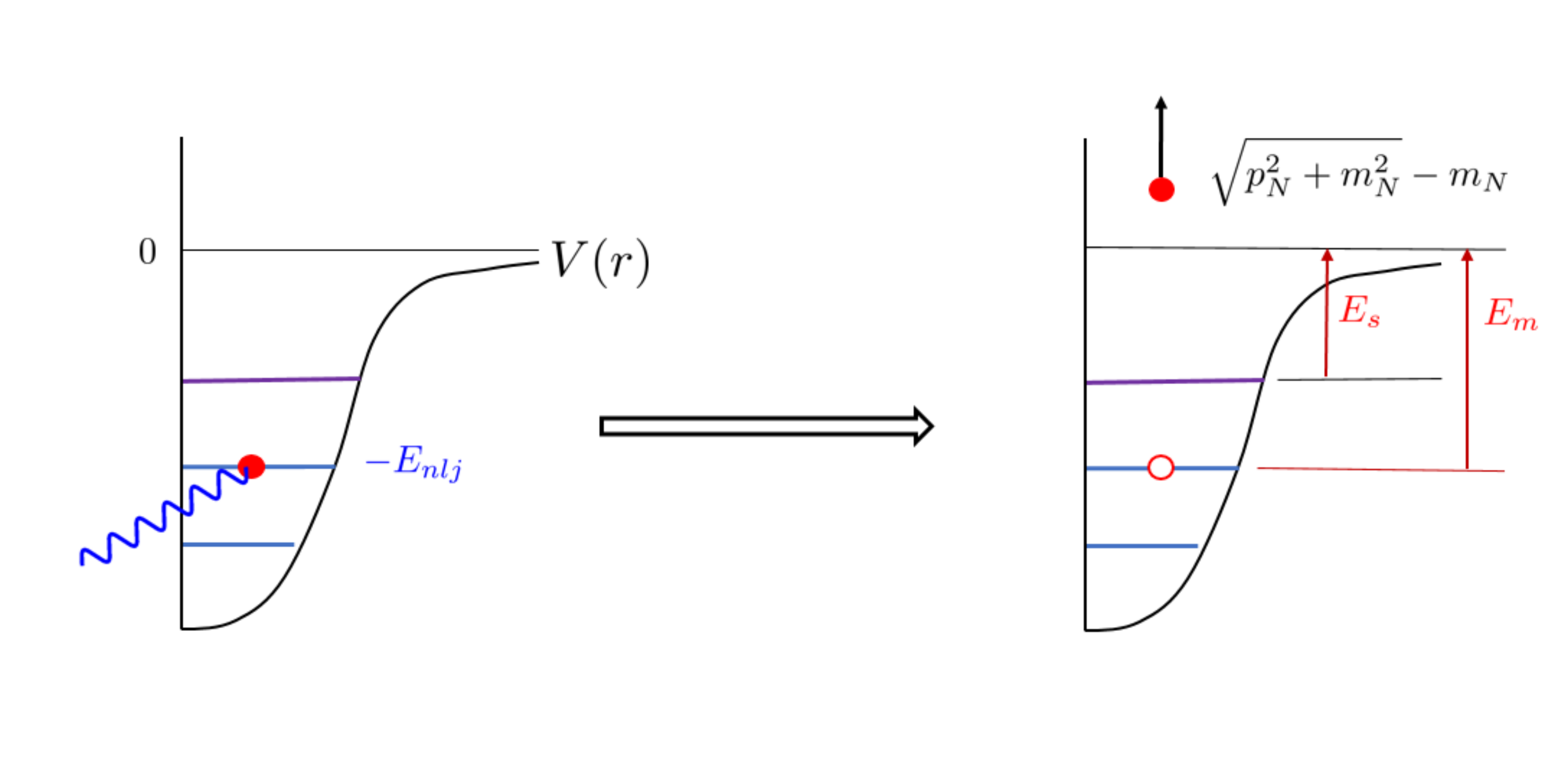}}
	\caption{(Color online) Schematic representation of an electroweak reaction within the independent particle shell model.}\label{fig:shell_model}
\end{figure}

Perhaps the simplest model is the independent particle shell model (IPSM), which consists of nucleons occupying discrete energy levels in a potential. The scattering process for this model is represented schematically by Fig.~\ref{fig:shell_model}. On the left is a representation of the interaction of a boson with a nucleon in a shell with energy $-E_{nlj}$, labeled with the usual quantum numbers $nlj$. The energy and momentum transfered to this nucleon results in producing an on-shell nucleon with relativistic kinetic energy $\sqrt{p_N^2+m_N^2}-m_N$ and leaving a hole in the residual nucleus. Energy conservation is then given by the delta function
\begin{equation}
\delta(k-\varepsilon'-E_{nlj}-\sqrt{p_N^2+m_N^2}+m_N)\,.
\end{equation}
Comparing this with Eq.~(\ref{eq:energy_conservation}) allows the identification of $\mathcal{E}=E_{nlj}-E_s$. Using the normalization conditions for the spectral function, the spectral function can the be identified as 
\begin{equation}
S_{SM}(p_m,E_s+\mathcal{E})=\sum_{n,l,j}(2j+1)n_{nlj}(p_m)
\delta(\mathcal{E}+E_s-E_{nlj})\,,\label{eq:spectral_IPSM}
\end{equation}
where $n_{nlj}(p_m)$ is the momentum distribution of a single nucleon in the $nlj$ shell and the factor $2j+1$ gives the number of neutrons in that shell. This spectral function then consists of set of delta-functions weighted by the total neutron momentum distribution for each shell.

Substituting Eq.~(\ref{eq:spectral_IPSM}) into Eq.~(\ref{eq:semi_simple}) gives the semi-inclusive cross section
\begin{align}
\frac{d\sigma}{dk'd\Omega_{k'}dp_Nd\Omega_N^L}=&\frac{G_F^2\cos^2\theta_cm_N{k'}^2p_N^2}{8(2\pi)^6 \varepsilon'E_N}\sum_{n,l,j}(2j+1)\frac{P(k_{0nlj})}{k_{0nlj}} v_0\widetilde{\mathcal{F}}^{2}_\chi n_{nlj}(p_m)\,,\label{eq:semi_IPSM}
\end{align}
where the neutrino momentum for each subshell is 
\begin{equation}
k_{0nlj}=\varepsilon'+E_N-m_N+E_{nlj}
\end{equation}
and the corresponding missing momentum is
\begin{align}
p_m=&\bigl[k_{0nlj}^2+{k'}^2+p_N^2-2k_{0nlj}k'\cos\theta_l-2k_{0nlj}p_N\cos\theta_N^L\nonumber\\
& +2k'p_N(\cos\theta_l\cos\theta_N^L+\sin\theta_l\sin\theta_N^L\cos\phi_N^L)\bigr]^\frac{1}{2}\,.
\end{align}

The inclusive cross section is
\begin{align}
\frac{d\sigma}{dk'd\Omega_{k'}}
=&\frac{G_F^2\cos^2\theta_cm_N{k'}^2}{8(2\pi)^5 \varepsilon'} \sum_{n,l,j}(2j+1)\int_0^\infty dk\frac{P(k)}{qk} \int_0^\infty dp_mp_mv_0\widetilde{\mathcal{F}}^2_\chi\nonumber\\
&\times n_{nlj}(p_m)
\delta(\mathcal{E}+E_s-E_{nlj})
\theta(p_m^+-p_m)\theta(p_m-p_m^-)\,,\label{eq:inclusive_IPSM}
\end{align}
where
\begin{equation}
p_m^+=\sqrt{(\omega-E_{nlj})(\omega-E_{nlj}+2m_N)}+q
\end{equation}
and	
\begin{equation}
p_m^-=\left|\sqrt{(\omega-E_{nlj})(\omega-E_{nlj}+2m_N)}-q\right|\,.
\end{equation}

\subsection{The Relativistic Fermi Gas Spectral Function (RFG)}\label{subsec:RFG}

\begin{figure}
	\centerline{\includegraphics[height=2.0in]{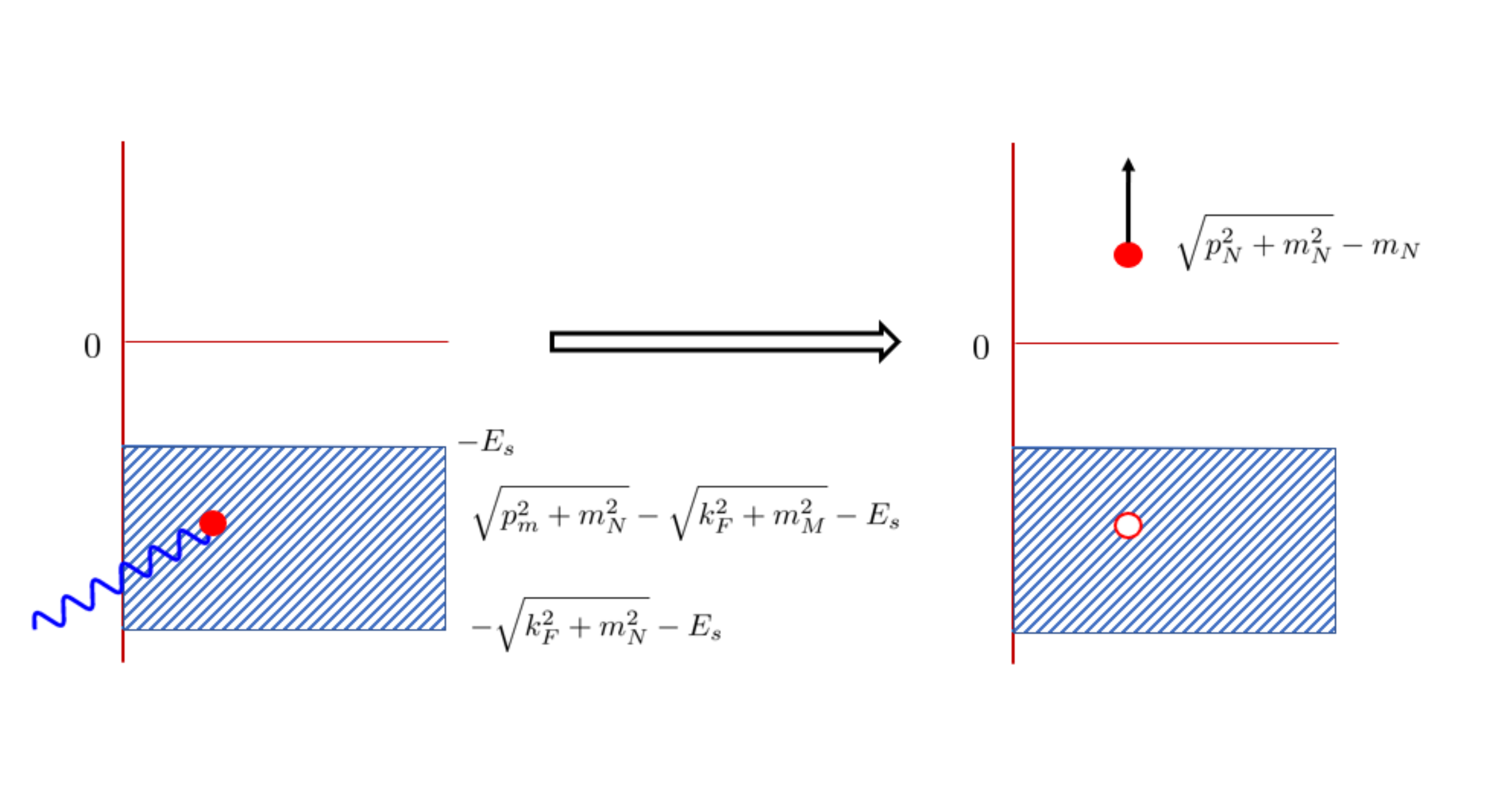}}
	\caption{(Color online) Schematic representation of an electroweak reaction within the relativistic Fermi gas model.}\label{fig:RFG}
\end{figure}

The relativistic Fermi gas model (RFG) was the model initially used in most event generators. It consists in describing the nucleus as an infinite gas of relativistic nucleons with all levels occupied up to a Fermi momentum $k_F$ and all levels above that empty. A schematic representation of the scattering process for this model is shown in Fig.~\ref{fig:RFG}. For consistency with other nuclear models we have chosen to shift the spectrum of the Fermi gas to negative energies such that we can place to top of the Fermi sea at the separation energy $-E_s$. The left part of this figure shows a boson striking a nucleon in the Fermi sea with energy $\sqrt{p_m^2+m_N^2}-\sqrt{k_F^2+m_N^2}-E_s$. On the right the final state with a nucleon with positive kinetic energy is produced leave a hole in the Fermi sea at the initial nucleon energy. Energy conservation is then given by the delta function
\begin{equation}
\delta(k-\varepsilon'+\sqrt{p_m^2+m_N^2}-\sqrt{k_F^2+m_N^2}-E_s-\sqrt{p_N^2+m_N^2}+m_N)\,.
\end{equation}
Comparing this with Eq.~(\ref{eq:energy_conservation}) allows the identification of $\mathcal{E}=\sqrt{k_F^2+m_N^2}-\sqrt{p_m^2+m_N^2}$. Using the normalization conditions for the spectral function, the spectral function can then be identified as 
\begin{equation}
S_{RFG}(p,\mathcal{E}+E_s,k_F)=\frac{3(2\pi)^3\mathcal{N}}{k_F^3}\delta(\mathcal{E}-\sqrt{k_F^2+m_N^2}+\sqrt{p_m^2+m_N^2})\theta(k_F-p_m)\,,\label{eq:S_RFG}
\end{equation}
where $\mathcal{N}$ is the number of neutrons. The spectral function is then a constant multiplied by a delta function located along the curve defined by $\mathcal{E}=\sqrt{k_F^2+m_N^2}-\sqrt{p_m^2+m_N^2}$ for $p_m<k_F$. This is in agreement with the derivation of the RFG spectral function in \cite{Barbaro:1998gu}.

Substituting Eq.~(\ref{eq:S_RFG}) into Eq.~(\ref{eq:semi_inclusive}) and then integrating over $\mathcal{E}$ with $\varepsilon=k$ gives the semi-inclusive cross section
\begin{align}
\frac{d\sigma}{dk'd\Omega_{k'}dp_Nd\Omega_N^L}=&\frac{3\mathcal{N}G_F^2\cos^2\theta_cm_N^2{k'}^2p_N^2}{8(2\pi)^3 \varepsilon'E_N k_F^3}\int_0^\infty dk\frac{P(k)}{k}\nonumber\\
&\times\int \frac{d^3p_m}{\sqrt{p_m^2+m_N^2}} v_0\widetilde{\mathcal{F}}^{2}_\chi \delta\left(\bm{k}-\bm{k}'-\bm{p}_N^L+\bm{p}_m\right)\nonumber\\
&\times \delta\left(k-\varepsilon'-E_s-E_N-T_F+\sqrt{p_m^2+m_N^2}\right)\nonumber\\
&\times\theta(k_F-p_m)\,.\label{eq:semi_RFG_0}
\end{align}
where
\begin{equation}
T_F=\sqrt{k_F^2+m_N^2}-m_N\,.
\end{equation}
Note that, since the nucleons in the Fermi sea have relativistic on-shell energies, it is necessary to introduce a factor of $\sqrt{p_m^2+m_N^2}^{-1}$ into the integral over $p_m$ and that the hole in the residual nucleus has momentum $-\bm{p}_m$.

Defining
\begin{equation}
\bm{p}_B=\bm{k}'+\bm{p}_N\,,
\end{equation}
\begin{equation}
E_B=\varepsilon'+E_s+T_F+E_N\,,
\end{equation}
\begin{equation}
\cos\theta_B=\frac{\bm{k}\cdot\bm{p}_B}{k p_B}\,.
\end{equation}
Integrating over $\bm{p}_m$ and rewriting the remain delta function as
\begin{align}
&\delta\left(k-\varepsilon'-E_s-T_F-E_N+\sqrt{(\bm{p}_B-\bm{k})^2+m_N^2}\right)\nonumber\\
&\qquad\qquad=\frac{\sqrt{(\bm{p}_B-\bm{k})^2+m_N^2}}{E_B-p_B \cos\theta_B }\delta(k-k_0)\,.
\end{align}
the semi-inclusive cross section becomes
\begin{align}
\left(\frac{d\sigma}{dk'd\Omega_{k'}dp_Nd\Omega_N^L}\right)_{k_F}=&\frac{3\mathcal{N}G_F^2\cos^2\theta_cm_N^2{k'}^2p_N^2}{8(2\pi)^3 \varepsilon'E_N k_F^3}\frac{ P(k_0)}{k_0}\frac{v_0\widetilde{\mathcal{F}}^{2}_\chi}{E_B-p_B \cos\theta_B} 
\theta(k_F-|\bm{k}-\bm{p}_B|)\,,\label{eq:semi_RFG}
\end{align}
where
\begin{equation}
k_0=\frac{E_B^2-p_B^2-m_N^2}{2(E_B-p_B\cos\theta_B)}\,,
\end{equation}
\begin{align}
p_m=|\bm{p}_B-\bm{k}_0|=\sqrt{q_0^2+p_N^2-2k_0p_N\cos\theta_N^L+2k'p_N(\cos\theta_l\cos\theta_N^L+\sin\theta_l\sin\theta_N^L\cos\phi_N^L)}
\end{align}
and
\begin{equation}
q_0^2=k_0^2-2 k'k_0\cos\theta_l+{k'}^2\,.
\end{equation}

The corresponding inclusive cross section can be obtained by integrating Eq.~(\ref{eq:semi_RFG_0}) $\bm{p}_N^L$ and then $d\omega_{p_m}$ yielding
\begin{align}
\left(\frac{d\sigma}{dk'd\Omega_{k'}}\right)_{k_F}=&\frac{3\mathcal{N}G_F^2\cos^2\theta_cm_N^2{k'}^2}{8(2\pi)^2 \varepsilon'k_F^3}\int_0^\infty dk\frac{P(k)}{qk} \int_0^\infty \frac{dp_mp_mv_0\widetilde{\mathcal{F}}^{2}_\chi}{\sqrt{p_m^2+m_N^2}}\nonumber\\
& \times \theta(k_F-p_m)\theta(p_+-p_m)\theta(p_m-p_-)\,,\label{eq:inclusive_RFG}
\end{align}
where
\begin{equation}
p_-=\left|\frac{\sqrt{(\omega-E_s-T_F)^2\xi(\xi+4m_N^2)}}{2\xi}-\frac{q}{2}\right|\,,
\end{equation}
\begin{equation}
p_+=\frac{\sqrt{(\omega-E_s-T_F)^2\xi(\xi+4m_N^2)}}{2\xi}+\frac{q}{2}
\end{equation}
and
\begin{equation}
\xi=q^2-(\omega-E_s-T_F)^2\,.
\end{equation}

\subsection{The Local Density Approximation (LDA) to the Spectral Function}\label{subsec:LDA}

An elaboration on the RFG is to include variation in the nuclear density by using the local density approximation (LDA). This assumes that RFG cross sections can be assumed to apply locally at each point in the nucleus. Starting with a representation of the nuclear density for a spherically symmetric nucleus $\rho(r)$ that satisfies the normalization condition
\begin{equation}
4\pi\int_0^\infty dr r^2 \rho(r)=\mathcal{N}\,.
\end{equation}
and the definition of nuclear density for the RFG 
\begin{equation}
\rho_{RFG}=\frac{\mathcal{N}}{V}=\frac{k_F^2}{3\pi^3}
\end{equation}
these can be equated to define a local Fermi momentum
\begin{equation}
\rho(r)=\frac{k_F^3(r)}{3\pi^2}
\end{equation}
giving
\begin{equation}
k_F(r)=\left( 3\pi^2\rho(r)\right)^\frac{1}{3}\,.
\end{equation}
Defining
\begin{equation}
\bar{\rho}(r)=\frac{\rho(r)}{\mathcal{N}}\,,
\end{equation}
the LDA spectral function can be defined as
\begin{equation}
S_{LDA}(p_m,\mathcal{E}+E_s)=4\pi\int_0^\infty 
dr r^2\bar{\rho}(r)S_{RFG}(p_m,\mathcal{E}+E_s,k_F(r))\,.
\end{equation}

The semi-inclusive and inclusive cross sections for this model can obtained from the RFG cross sections in Eqs.~ (\ref{eq:semi_RFG}) and (\ref{eq:inclusive_RFG}) by a method similar to that used to produce the spectral function. The semi-inclusive cross section is
\begin{equation}
\frac{d\sigma}{dk'd\Omega_{k'}dp_Nd\Omega_N^L}=4\pi\int_0^\infty dr r^2\bar{\rho}(r) \left(\frac{d\sigma}{dk'd\Omega_{k'}dp_Nd\Omega_N^L}\right)_{k_F(r)}
\end{equation}
and the inclusive cross section is
\begin{equation}
\frac{d\sigma}{dk'd\Omega_{k'}}=4\pi\int_0^\infty dr r^2\bar{\rho}(r)\left(\frac{d\sigma}{dk'd\Omega_{k'}}\right)_{k_F(r)}\,.
\end{equation}

\subsection{A More Realistic Spectral Function (Rome)}\label{subsec:Rome}

More realistic spectral functions can be constructed using $(e,e'p)$ data for low missing energies and models of the interacting nuclear system for larger values of missing energy and missing mass.
As and example of this approach we use a spectral function for $^{16}\mathrm{O}$  constructed by Benhar, et al. \cite{Benhar:1994hw,Benhar:2005dj}, which contains information obtained from $(e,e'p)$ measurements for relatively small values of $p_m$ and $E_m$, providing information about the shell structure of the nucleus. This can be reproduced by means of an IPSM approximation with the individual shells artificially widened using Lorentzians, together with a representation of contributions from correlations which in the model of the Rome group are calculated in nuclear matter. These are then produced for finite nuclei by means of the LDA. The resulting spectral function is of the form
\begin{equation}
S_{Rome}(p_m,E_m)=S_{IPSM(p_m,E_m)}+S_{corr}(p_m,E_m)\,.
\end{equation}
Any constants need to combine the two combinations are adjusted so that the total satisfies the normalization condition in Eq.~ (\ref{eq:spectral_norm}). This, and similar approaches, represent the current state of the art. We will use this spectral function, which we will refer to as the Rome spectral function, as a bench-mark against which the other models used here will be compared.

\section{Results}\label{sec:results}

\subsection{Spectral Functions}

We now proceed with a discussion of several types of spectral functions of varying degrees of sophistication --- all of the spectral functions and cross sections shown here are for $^{16}\mathrm{O}$. We consider four models for the spectral functions beginning with a simple independent-particle shell model with relativistic mean field single-particle wave functions (IPSM-RMF) which captures the basic essentials of the nuclear shell structure of a nucleus such as $^{16}$O. This is followed by going to the other extreme and discussing the relativistic Fermi gas (RFG) model which is designed to contain only the basic properties of infinite nuclear matter; it is, in fact, a model where $A \rightarrow \infty$ and the only aspect of finite nuclei it contains is a scale, the Fermi momentum $k_F$. This is included here despite its simplicity (as we shall see, too simple for semi-inclusive studies) since it forms the basis for many of the event generators currently being employed. Attempts have been made to improve on the extreme RFG model by incorporating a density-dependent Fermi momentum that follows the ground-state density of a given nucleus, the so-called local density approximation (LDA), and this provides the third model in the present study. These simplified approaches are then compared with a state of the art spectral function obtained by the Rome group. In the following sections we proceed to obtain the inclusive and semi-inclusive cross sections using the four models, and there we find that the former do not differ significantly, although when a nucleon is presumed to be detected (sem-inclusive reactions), the resulting cross section are strongly dependent on the level of sophistication contained in the various models.

\subsubsection{IPSM-RMF Spectral Function}

\begin{figure}
	\centerline{\includegraphics[height=2.5in]{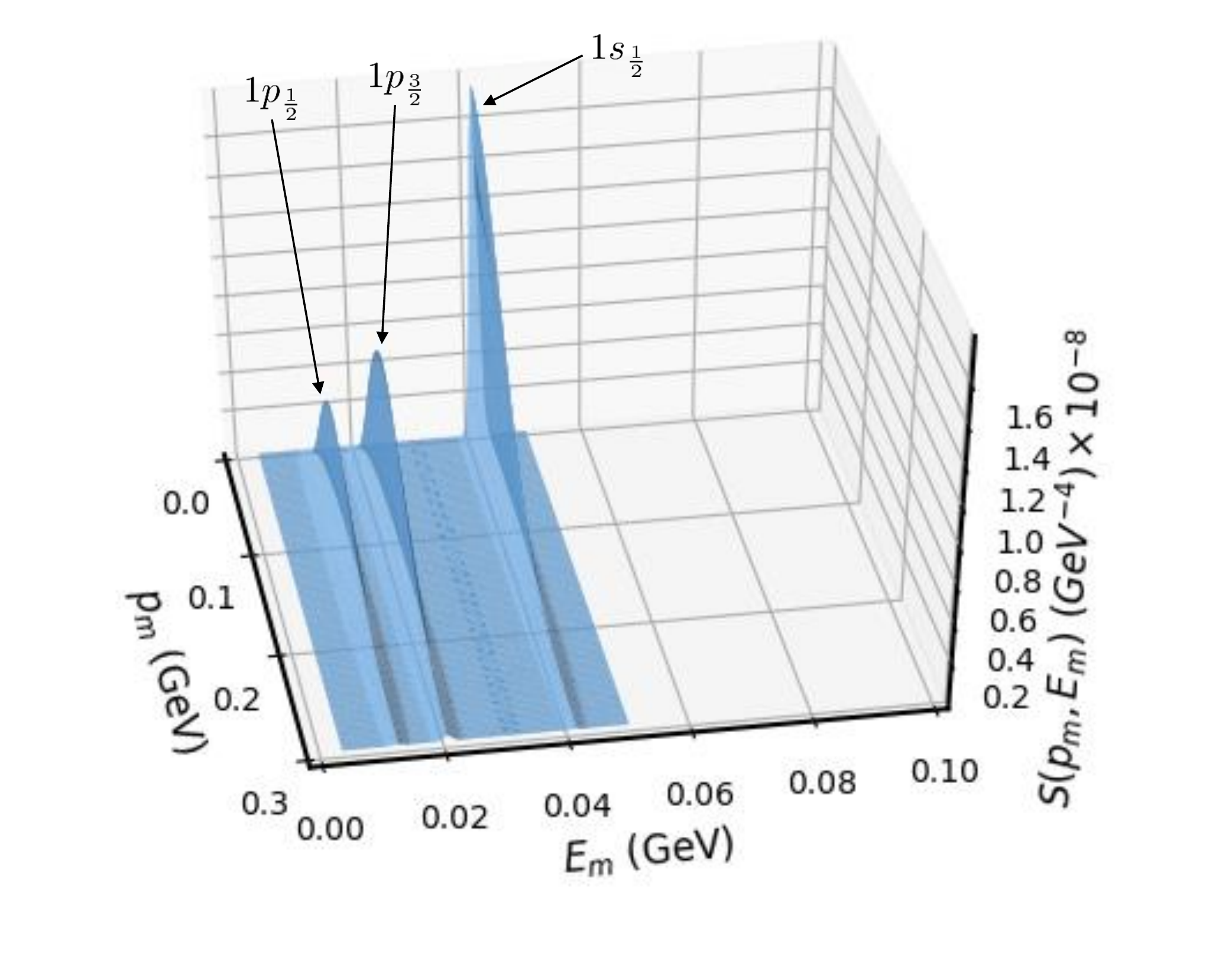}}
	\caption{(Color online) Independent particle shell model spectral function for $^{16}$O using RMF wave functions \cite{Horowitz:1981xw} for the hole states.}\label{fig:spectral_function_RMF}
\end{figure}

An example of IPSM spectral functions is presented in Fig.~\ref{fig:spectral_function_RMF}. This uses the relativistic mean field model (RMF) of Horowitz and Serot \cite{Horowitz:1981xw} for $^{16}\mathrm{O}$ to obtain the wave functions for the shells occupied by neutrons. In the case of a model such as this, that produces wave functions in the form of Dirac spinors it is necessary to project the wave functions onto positive-energy plane-wave Dirac spinors and renormalize to obtain factorizable cross sections. We refer to this as the IPSM-RMF model. In this figure the delta functions have been replaced by Gaussians to facilitate visualization. The location of each is indicated in the this figure.

\subsubsection{RFG Spectral Function}

\begin{figure}
	\centerline{\includegraphics[height=2.5in]{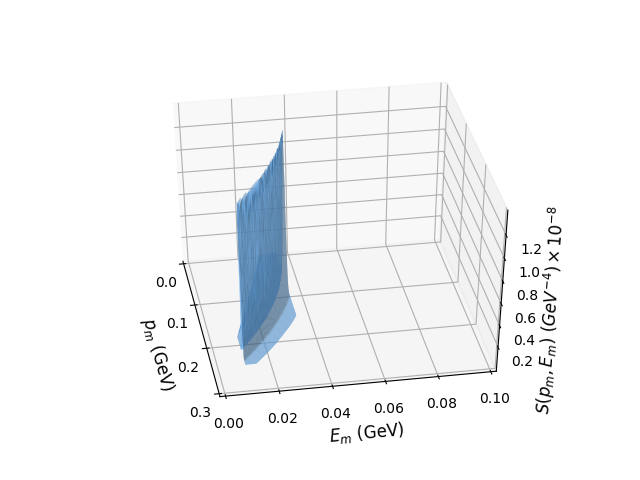}}
	\caption{(Color online) Relativistic Fermi gas spectral function for $^{16}$O using $k_F=230$ MeV/c.}\label{fig:spectral_RFG}
\end{figure}

The RFG spectral function is represented by Fig.~\ref{fig:spectral_RFG} for $k_F=230$ MeV, which is a reasonable choice for $^{16}\mathrm{O}$ \cite{Megias:2017cuh}. The delta function has again been replaced by a Gaussian to facilitate visualization.
Clearly, even when compared with the simple IPSM (see Fig.~\ref{fig:spectral_function_RMF}), the RFG spectral function in Fig.~\ref{fig:spectral_RFG} is a rather crude approximation in which the shell model structure is completely gone and replaced with a $\delta$-function ``wall''. This should be expected since the RFG is the $A\rightarrow \infty$ limit and cannot be expected to represent the detailed level structure of a nucleus as light as oxygen.

\subsubsection{LDA Spectral Function}

\begin{figure}
	\centerline{\includegraphics[height=2.5in]{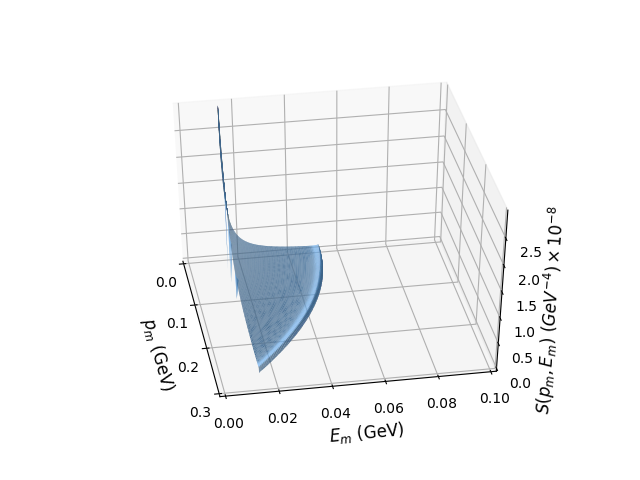}}
	\caption{(Color online) Spectral function in the local density approximation The coordinate space $\rho(r)$ is obtained from a three-parameter Fermi function fit to the proton distribution for $^{16}$O.}\label{fig:spectral_LD}
\end{figure}

\begin{figure}
	\centerline{\includegraphics[height=2.5in]{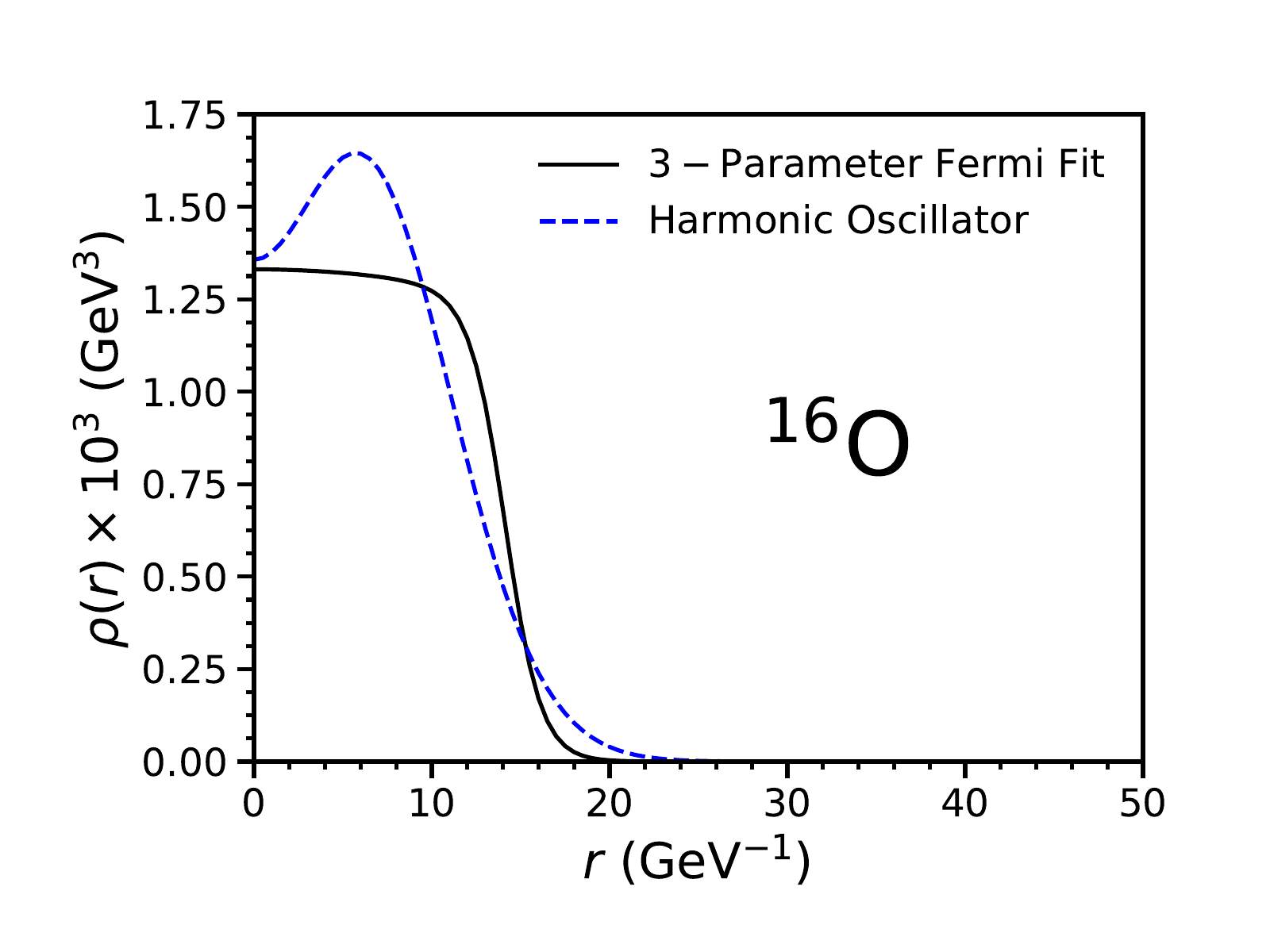}\includegraphics[height=2.5in]{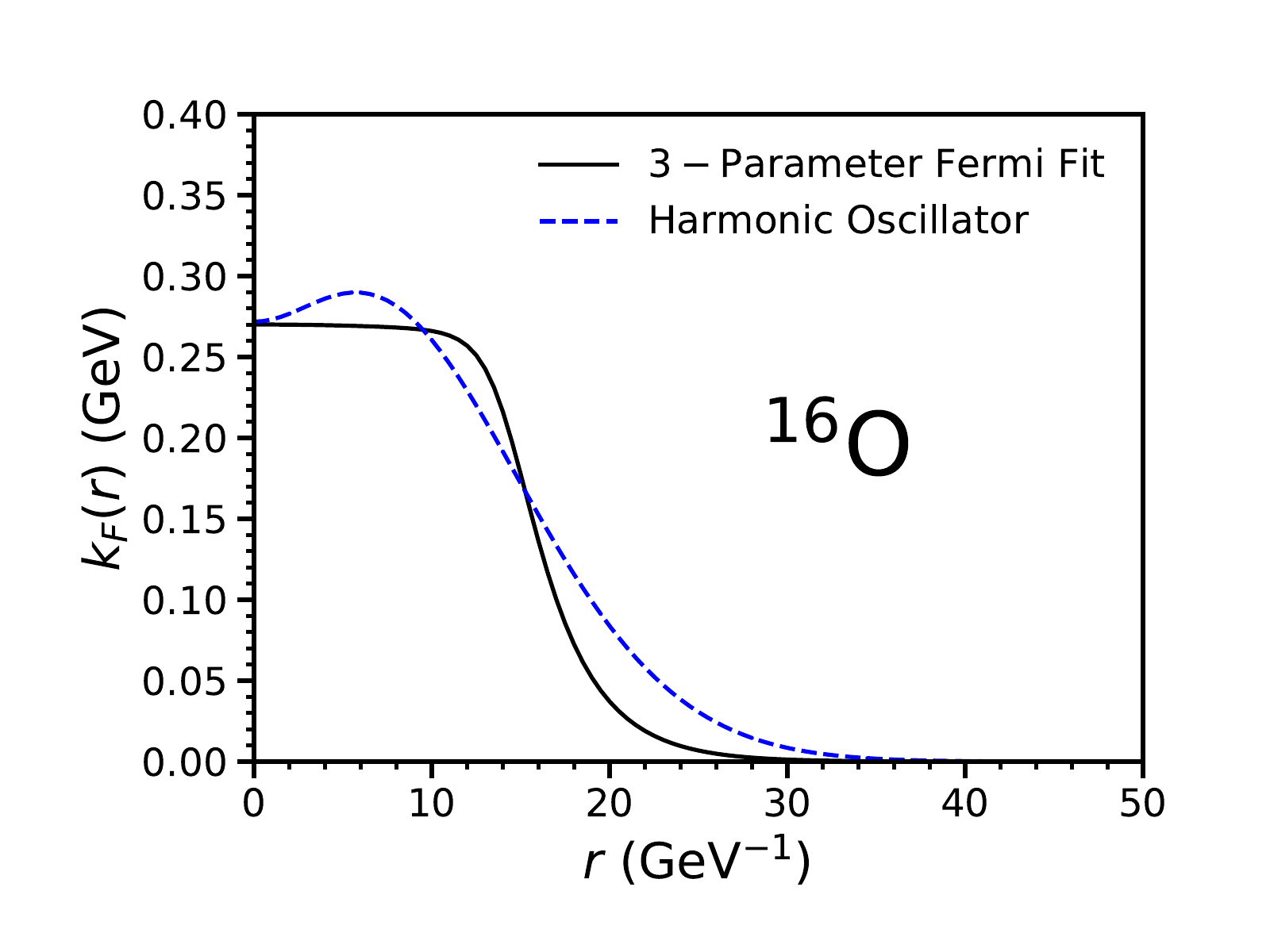}}
	\caption{(Color online) Density $\rho(r)$ and density-dependent Fermi momentum $k_F(r)$ versus $r$ used in the LDA for $^{16}$O. Results are shown using a three-parameter Fermi fit to the proton density and the density obtained from a simple harmonic oscillator (HO) model. }\label{fig:r_densities}
\end{figure}

In an attempt to embed some aspects of the structure of the nuclear ground state into a model that still retains some of the simplicity of the RFG an extension to the LDA can be employed.  Figure \ref{fig:spectral_LD} shows the LDA spectral function obtained from a three-parameter Fermi fit \cite{DeJager:1987qc} to the $^{16}\mathrm{O}$ charge density. Note that this is singular for $\mathcal{E}=0$: this singularity is associated with nuclear densities for larger values of $r$ where the density and the local Fermi momentum and its derivative become small and consequently the local spectral function becomes large. This is illustrated by Fig.~\ref{fig:r_densities} that shows plots of the $\rho(r)$ and $k_F(r)$ for $^{16}\mathrm{O}$ obtained from the 3-parameter Fermi fit and from a harmonic oscillator model. While it may have been hoped that the LDA would be a significant improvement over the RFG model for the spectral function, it is clear from Fig.~\ref{fig:spectral_LD} that the LDA yields results for the spectral function (and for the semi-inclusive cross sections; see Sect.~IV.D) that are quite different in form to the IPSM-RMF results in Fig.~\ref{fig:spectral_function_RMF}.

\subsubsection{Rome Spectral Function}

\begin{figure}
	\centerline{\includegraphics[height=2.5in]{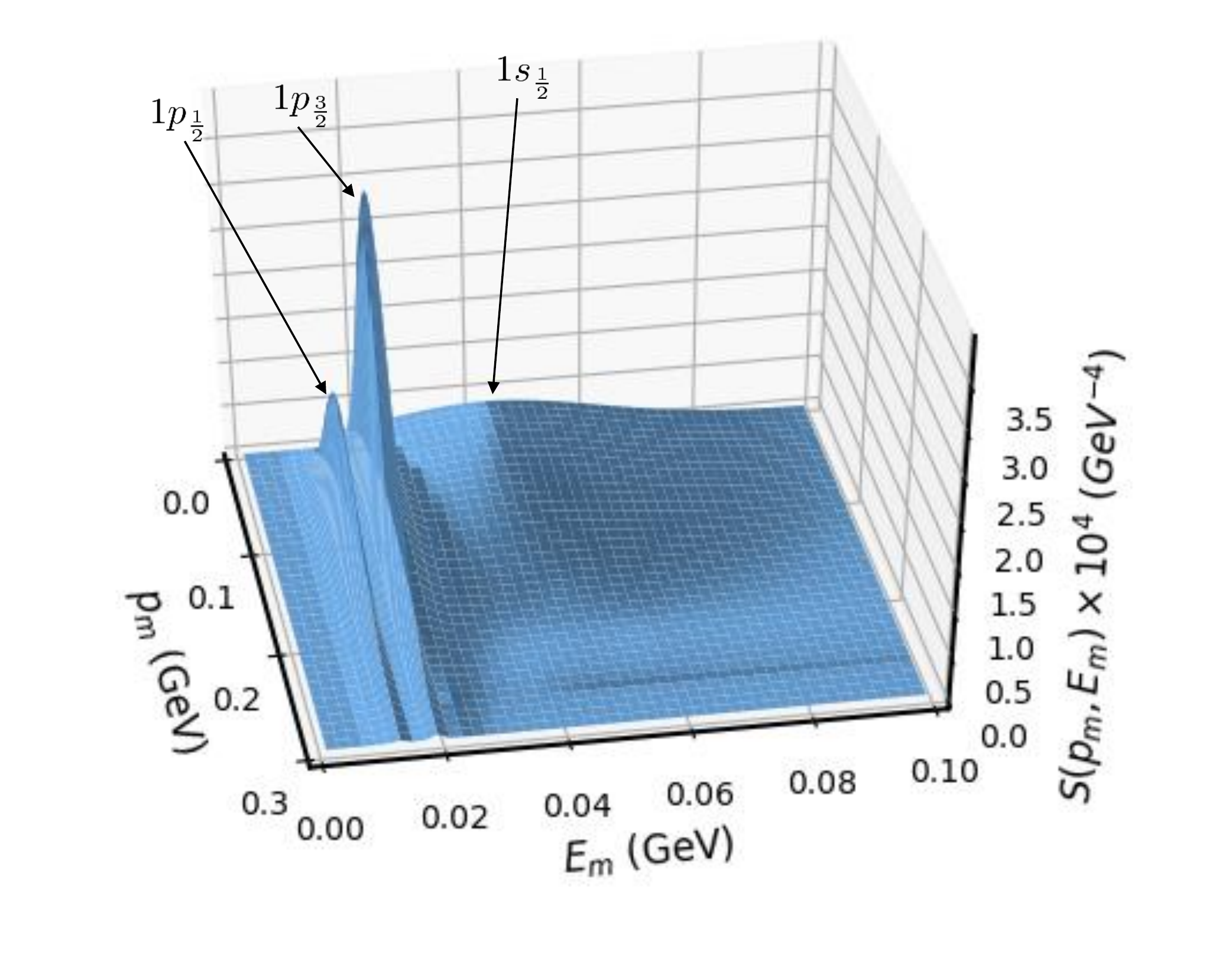}\includegraphics[height=2.5in]{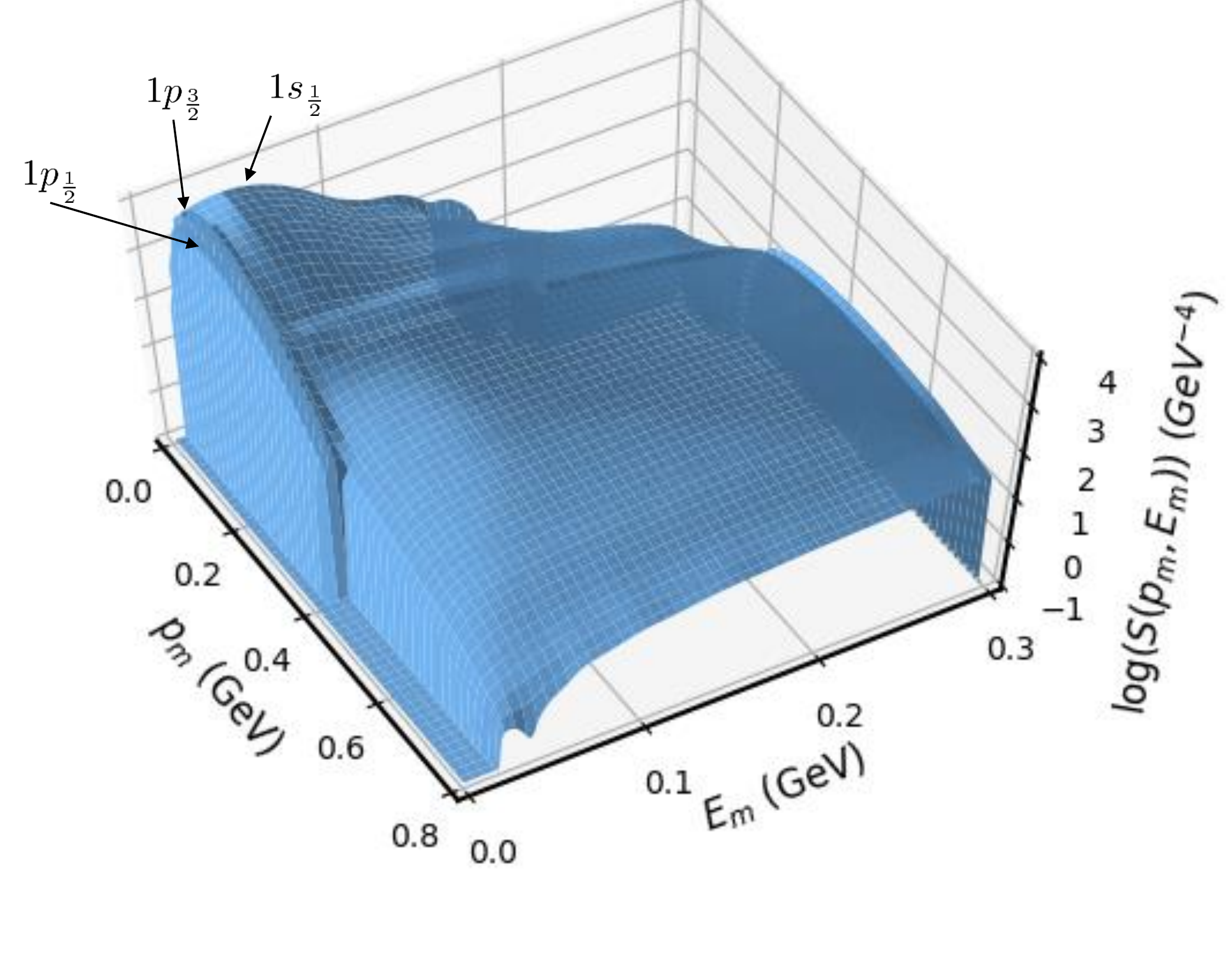}}
	\caption{(Color online) The Rome spectral function \cite{Benhar:1994hw,Benhar:2005dj}.}\label{fig:spectral_benhar}
\end{figure}

The Rome spectral function is displayed in Fig.~\ref{fig:spectral_benhar}. The linear plot of the left-hand panel clearly displays the contribution of the shell structure of the nucleus. The location of the IPSM shells used here is idicated. A particularly noticeable feature is the substantial broadening of the $1s_\frac{1}{2}$ contribution which occurs since the energy necessary to remove a nucleon from this shell is greater than that required to remove two nucleons from the p shells. This results in substantial mixing of contributions of these shells with the two-nucleon continuum via the interaction of the nucleons resulting in the substantial broadening of the $1s_\frac{1}{2}$ shell. The  $1p_\frac{3}{2}$ and  $1p_\frac{1}{2}$ shells a given a smaller width to model details of nuclear structure at lower missing energies. The contribution of the correlation part of the spectral function is more clearly displayed by the semi-log plot in the right-hand panel. Note the presence of a ridge in $E_m$ running across the spectral function at an almost constant value of $E_m\sim 2.5\ \mathrm{GeV}$.

A rough indication of the properties of the RFG, LDA, Rome and IPSM-RMF models is provided by Table~\ref{tab:occupation}. Here, the occupation number for each model is calculated over five ranges of $E_m$. The first three regions are chosen to roughly correspond to the contributions of the $1p_\frac{1}{2}$, $1p_\frac{3}{2}$ and $1s_\frac{1}{2}$ shells of the Rome and IPSM-RMF spectral functions. It is clear that the pattern of occupation of the first three regions for the Rome and IPSM-RMF models are similar, but indicate the redistribution of strength in the Rome spectral function due to the interactions of the nucleons. Note also that, of the 8 neutrons in $^{16}$O, only 0.8 show up at missing energies greater than 0.1 GeV for the Rome model. The pattern of distribution of strength for RFG and LDA spectral functions is very different from those of the other two models, as is obvious from  Fig.~\ref{fig:spectral_RFG} and Fig.~\ref{fig:spectral_LD}. 

\begin{table}
	\begin{tabular}{|c|c|c|c|c|c|}\hline\hline
		$E_m$ (GeV) & RFG & LDA & Rome & IPSM & shell\\ \hline
		0.000--0.0165 & 3.04  & 4.22 & 1.51 & 2.00 & $1p_\frac{1}{2}$ \\
		0.0165--0.025 & 2.78 & 2.28 & 3.47 & 4.00	& $1p_\frac{3}{2}$ \\
		0.025--0.100 & 2.18 & 1.50 & 2.22 & 2.00 & $1s_\frac{1}{2}$ \\
		0.100--0.200 &  &  & 0.60 &   & \\
		0.200--0.300 &  &  & 0.20 &   & \\\hline\hline	
	\end{tabular}
	\caption{This table shows the neutron occupation number associated with five ranges of missing energy $E_m$ for the RFG, LDA, Rome and IPSM spectral functions. The designated shell of the IPSM is given in the last column. \label{tab:occupation}}
\end{table}

\subsection{Momentum Density Distributions}\label{subsec:momdist}

\begin{figure}
 	\centerline{\includegraphics[height=2.5in]{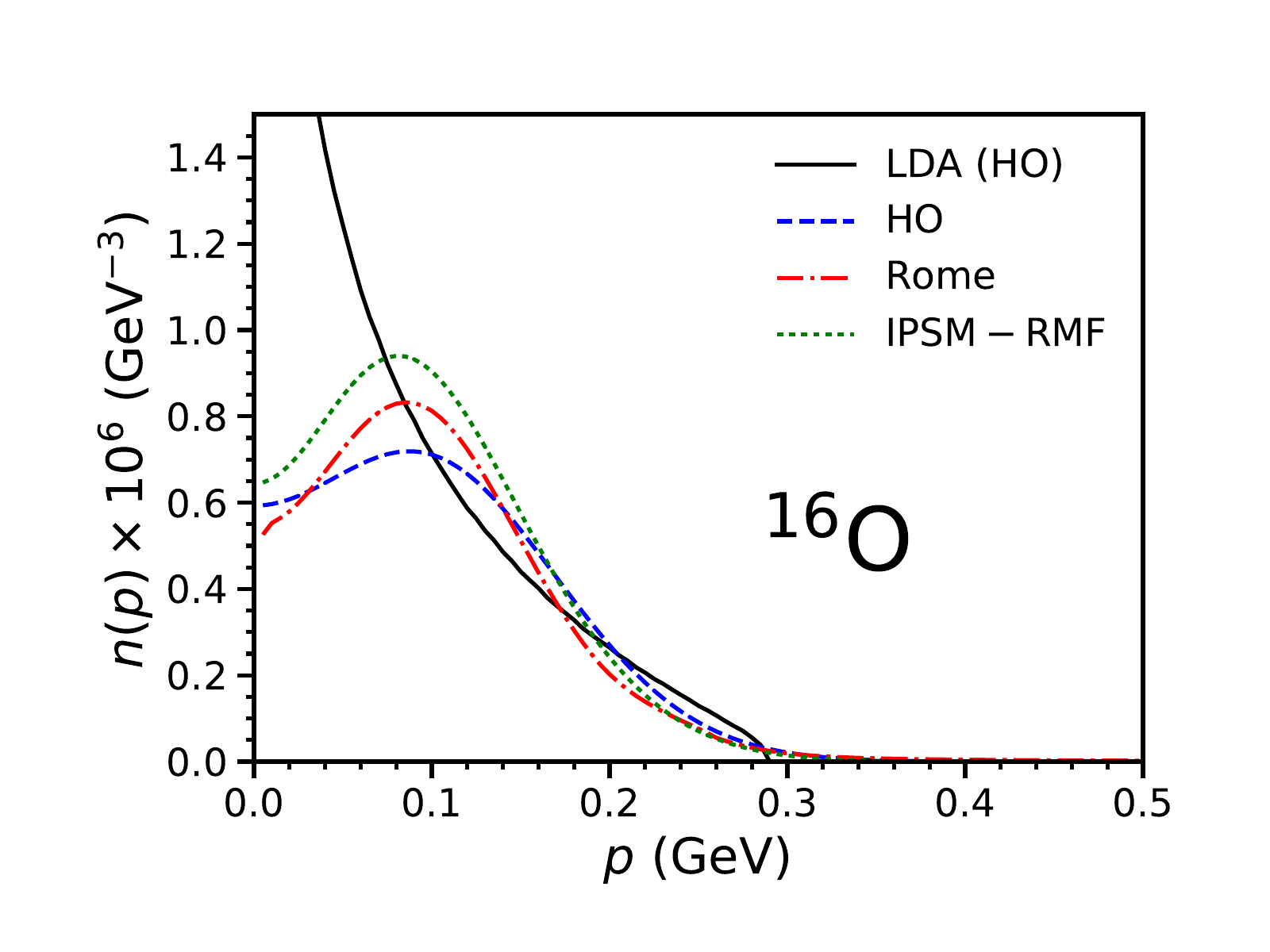}\includegraphics[height=2.5in]{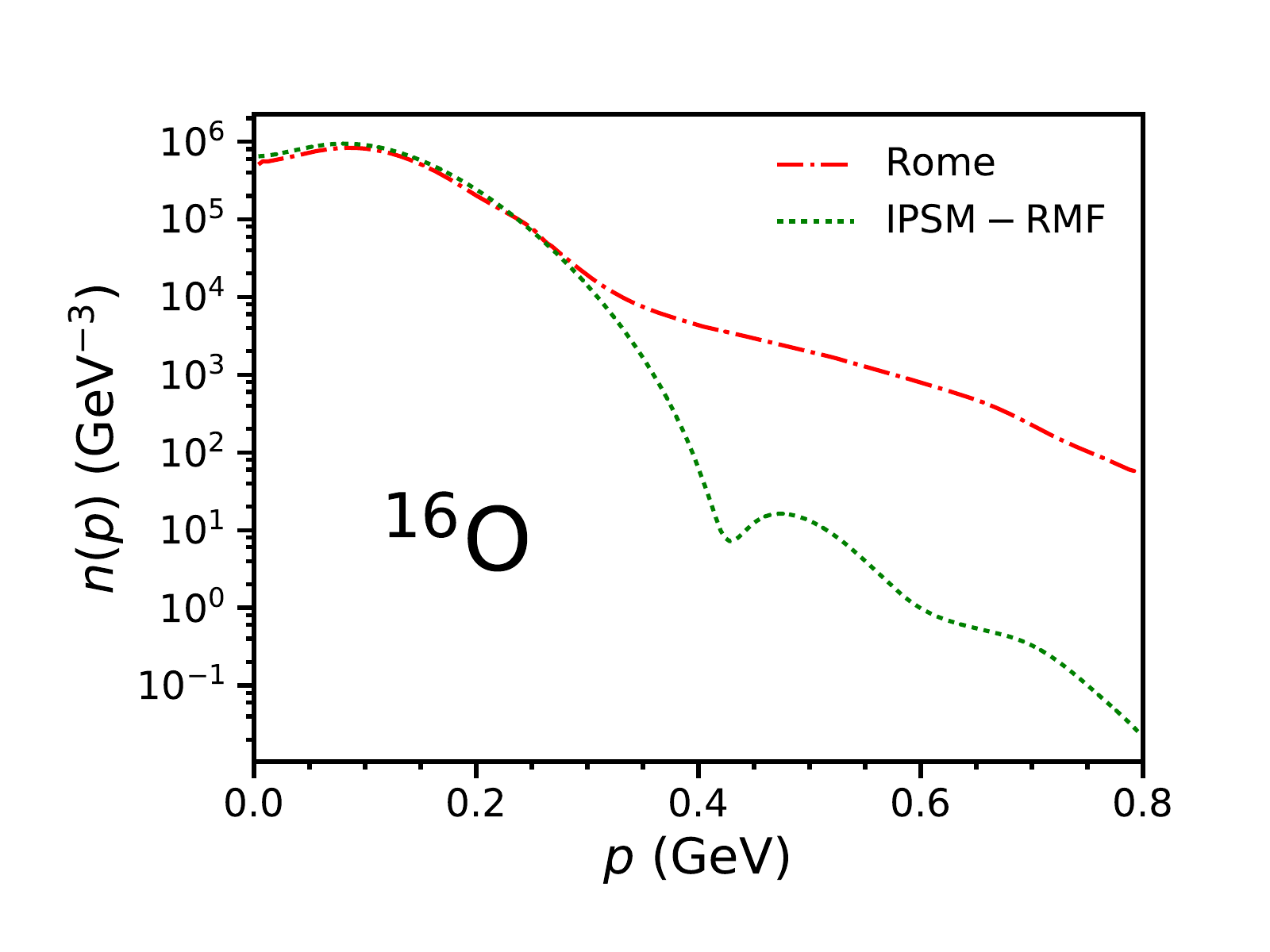}}
	\caption{(Color online) Momentum density distributions $n(p)$ versus $p$ obtained from LDA spectral function using the HO density, the density distributions calculated directly from the momentum-space oscillator wave functions, the distribution obtained from the Rome spectral function and the distribution using the momentum-space wave functions for the IPSM-RSM are shown in a linear plot on the left. A semilog plot of the Rome and IPSM-RSM distributions over an extended momentum range are on the right.}\label{fig:p_densities}
\end{figure}

Equation (\ref{eq:spectral_momentum}) requires that it should be possible to determine the nuclear momentum density distribution by integrating the spectral function over the missing energy.  This provides an interesting test for the LDA spectral function in the case where the momentum distribution can be calculated directly from the square of the momentum space wave functions as well as from the spectral function. An example of this would be an independent particle model such as a simple harmonic oscillator (HO) model. The squares of the coordinate-space wave functions can be used to provide a coordinate-space nuclear density distribution that can be used to produce an LDA spectral function and the squares of the momentum-space wave functions can be used to directly produce the momentum density distribution. The result of using these two approaches is shown in Fig.~\ref{fig:p_densities}. This the left-hand panel of this figure shows the momentum distributions calculated directly from the harmonic oscillator wave functions as well as from  Eq.~(\ref{eq:spectral_momentum}) for the LDA spectral function using the harmonic oscillator coordinate-space nucleon distribution.
Note that the momentum density distribution derived from the local density spectral function does not reproduce the direct calculation using the harmonic oscillator wave functions. It is in fact singular at $p=0$, although it still satisfies the normalization integral in Eq.~(\ref{eq:spectral_norm}). On the other hand, the momentum distribution obtained from the realistic spectral function is reasonable. A plot of the the momentum distribution for the IPSM-RFM is included for completeness. This indicates that great care is necessary when using the LDA. The right-hand panel is a semi-log plot of the momentum distributions for the Rome and IPSM-RFM spectral function and shows the effect of short-range correlations in the Rome spectral function. Importantly, note that the high-$p$ tail of the Rome momentum distribution arises entirely from large values of $E_m$, with very little coming from the valence structure at low $E_m$. In discussion semi-inclusive cross sections below we shall see that the valence region typically is completely dominant and thus that this hight-$p$ tail does not play much of a role, at least where neutrino cross sections are relatively large.

\subsection{Inclusive Cross Sections}\label{subsec:incl}

\begin{figure}
	\centerline{\includegraphics[height=2.5in]{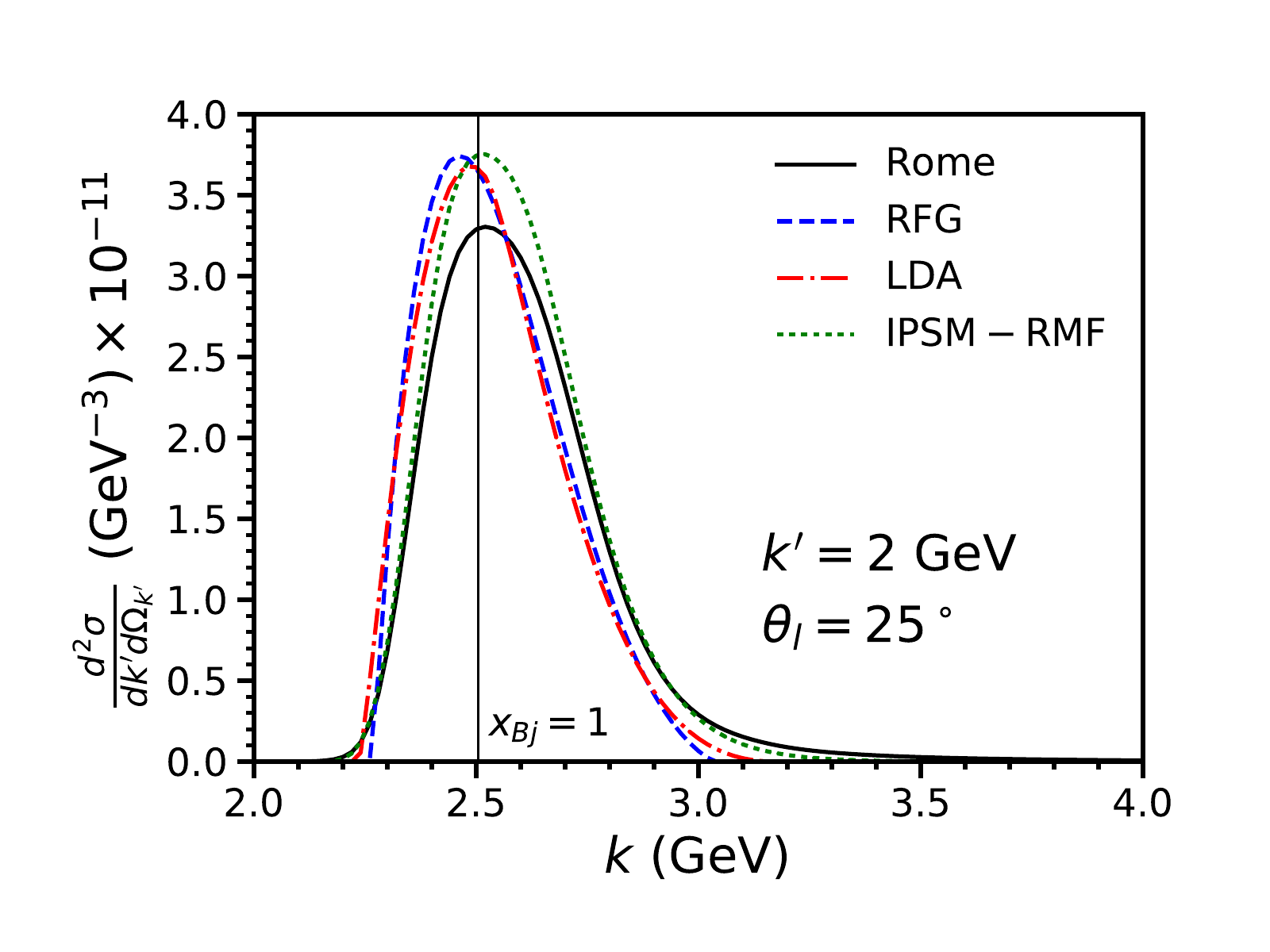}}
	\caption{(Color online) Inclusive cross sections at $k'=2\ \mathrm{GeV}$ and $\theta_l=25^\circ$ for the four models discussed in the text. The vertical line gives the value of $k$ corresponding to $x_{Bj}=1$ for this kinematics.}\label{fig:inclusive}
\end{figure}

Figure \ref{fig:inclusive} shows the inclusive cross sections for the various models for muon momentum $k'=2\ \mathrm{GeV}$ and $\theta_l=25^\circ$ without performing the weighted integral over the initial neutrino momentum $k$.
All of the cross sections are of similar shape with the peaks of the RFG and LDA cross sections occuring at approximately the same values of $k$ while the peaks from the realistic spectral function and the IPSM-RMF occur at somewhat higher values. The positions of the peaks can be adjusted to match data by introducing an additional phenomenological shift to the calculations relying on the RFG, while adjustment of the position of the spectral function peak can be moved by including a factorized final-state interaction function \cite{Benhar:2005dj}. However, if the models are to be used as input to event generators, no final-state interaction corrections should be included since these effects are supposedly produced by the event generator.

The smaller size of the peak for the Rome model reflects the shifting of strength due to nuclear interactions which also manifests itself in the slower falloff of the tail above approximately 3 GeV as compare to the IPSM-RMF result. If the IPSM-RMF result is multiplied is multiplied by 0.88, the peak corresponds in position and value to that of the Rome result for $k<2.6\ \mathrm{GeV}$, with the Rome result becoming larger above this point. This becomes appreciable at approximately $3\ \mathrm{GeV}$. 

\begin{table}
	\begin{tabular}{|l|c|c|}\hline\hline
		 & $k\ \mathrm{(GeV)}$ & cross section \\ \hline
	RFG & 2.463 &	$3.7440\times 10^{-11}$	    \\
	LDA & 2.488	 & $3.6814\times 10^{-11}$	  \\
	IPSM-RMF & 2.516 &	$3.7536\times 10^{-11}$	  \\
	Rome & 2.521 &	$3.3049\times 10^{-11}$	 \\ \hline	 
	\end{tabular}
	\caption{Positions of the peaks in the inclusive cross sections for the four models at $k'=2.0\ \mathrm{GeV}$ and $\theta_l=25^\circ$. The cross section is in units of $\mathrm{GeV}^{-3}\mathrm{sr}^{-1}$.}\label{tab:incl_maxima}
\end{table}

\subsection{Semi-Inclusive Cross Sections}\label{subsec:semi}

The purpose of the present study is to provide some insight into how various models may appear similar when integrated results provide the focus ({\it i.e.,} for inclusive cross sections), but are in fact quite different when more differential responses provide the focus ({\it i.e.,} semi-inclusive cross sections in this work). Although we are in a position to do a systematic study for a wide range of kinematics, that is not our purpose at present. Instead we have chosen two representative choices of kinematics, the first (I) being where the neutrino cross sections are at their largest and the second (II) being a choice where the cross sections are still within an order of magnitude of their maximum, but display rather different features that can be traced back to the underlying spectral functions discussed above.

\begin{figure}
	\centerline{\includegraphics[height=2.5in]{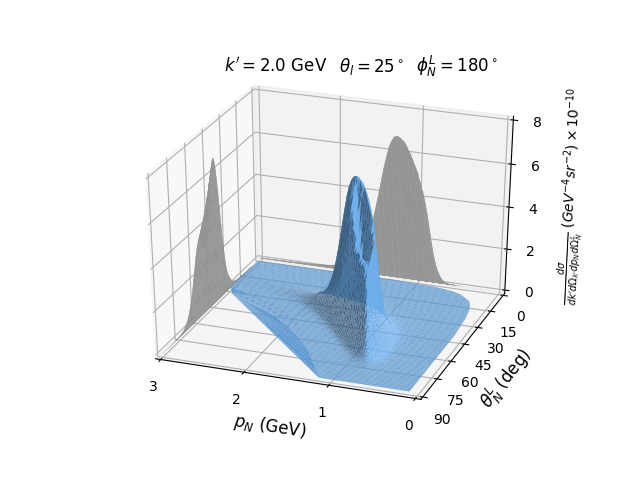}\includegraphics[height=2.5in]{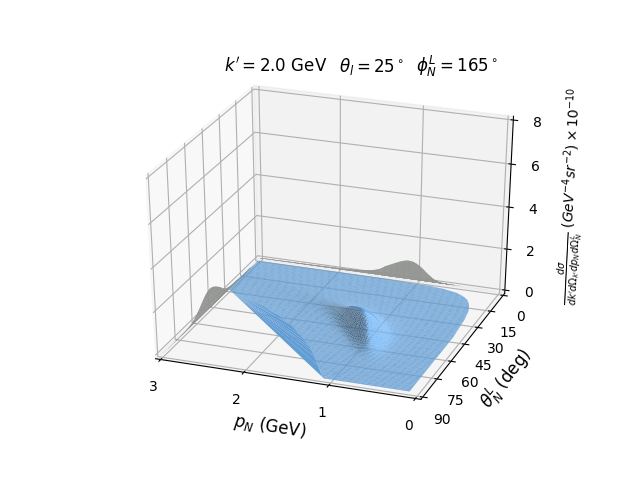}}
	\caption{(Color online) The semi-inclusive cross section for $k'=2\ \mathrm{(GeV)}$, $\theta_l=25^\circ$ for $\phi_N^L=180^\circ$ (case I) and $\phi_N^L=165^\circ$ (case II) using the Rome spectral function.}\label{fig:semi_spectral}
\end{figure}

We begin by considering the semi-inclusive cross sections for the various models at muon momentum $k'=2\ \mathrm{GeV}$, lepton scattering angle $\theta_l=25^\circ$ and nucleon azimuthal angles $\phi_N^L=180^\circ$ (case I) and $\phi_N^L=165^\circ$ (case II) as a function the detected nucleon moment $p_N$ and angle  $\theta_N^L$. The cross sections for the Rome spectral function is shown in Fig.~\ref{fig:semi_spectral}. For case I, the peak of the cross section is located at $p_N=1.29\ \mathrm{GeV}$ and $\theta_N^L=41.4^\circ$ with a value of $6.74\times 10^{-10}\ \mathrm{GeV}^{-4}\,\mathrm{sr}^{-2}$. For clarity, the cross section is projected on the side and back planes.It should be noted that, at the peak of the cross section, all but $0.2\%$ of the cross section comes from $p_m<0.12\ \mathrm{GeV})$ and $E_m<0.1\ \mathrm{GeV}$. The shape is not simple with a hollow appearing at the upper side of the peak. This is due to the features in the Rome spectral function associated with shell structure of the nucleus. For case II, where all of the contributions of the shell-model features to the cross section are similar in size, the shape of the cross section is relatively simple.

\begin{figure}
	\centerline{\includegraphics[height=2.5in]{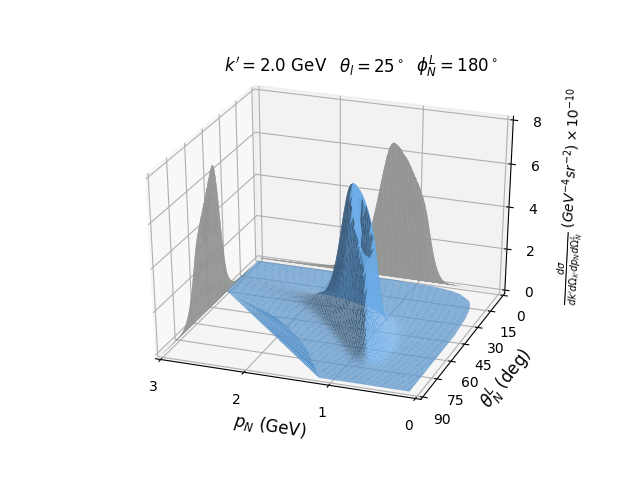}\includegraphics[height=2.5in]{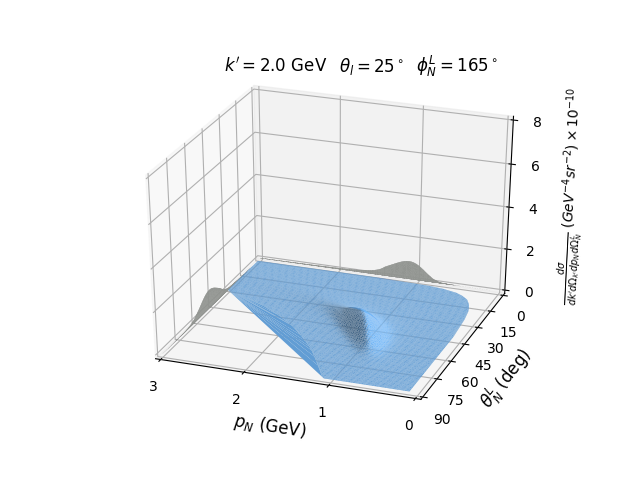}}	\caption{(Color online) The semi-inclusive cross section for $k'=2\ \mathrm{(GeV)}$, $\theta_l=25^\circ$ for $\phi_N^L=180^\circ$ (case I) and $\phi_N^L=165^\circ$ (case II) using the Rome spectral function. While in Fig.~\ref{fig:semi_spectral} all contributions to the cross section have been included, here contributions from the interference response functions have been omitted.}\label{fig:semi_spectral_star}
\end{figure}

It is important to notice that the effective single-nucleon cross section given in Appendix \ref{sec:appA} has interference response functions that are multiplied by either $\cos\phi_N$ or $\cos 2\phi_N$ where $\phi_N$ is the azimuthal angle of the detected proton about the direction of the three-momentum transfer $\bm{q}$ rather than $\phi_N^L$ which is the azimuthal angle about the neutrino beam direction. These interference contributions are often neglected in simple models. The importance of these contributions is indicated by Fig.~\ref{fig:semi_spectral_star}  where the previous calculation is repeated but with all of these interference contributions set to zero. Comparison with Fig.~\ref{fig:semi_spectral} shows that the maximum value of the cross section is decreased and that there is a slight change in shape. 

\begin{figure}
	\centerline{\includegraphics[height=2.5in]{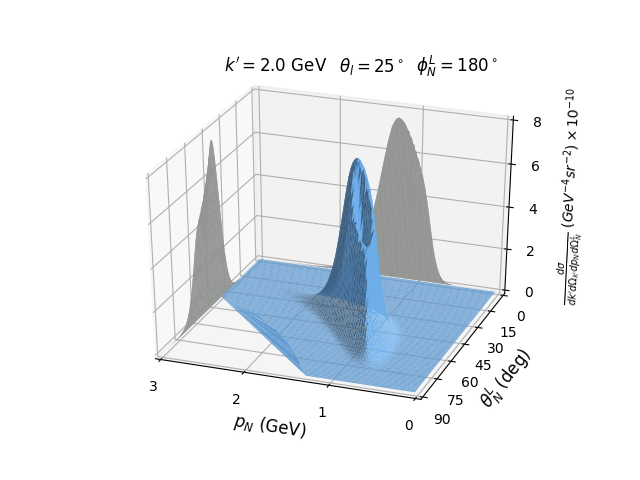}\includegraphics[height=2.5in]{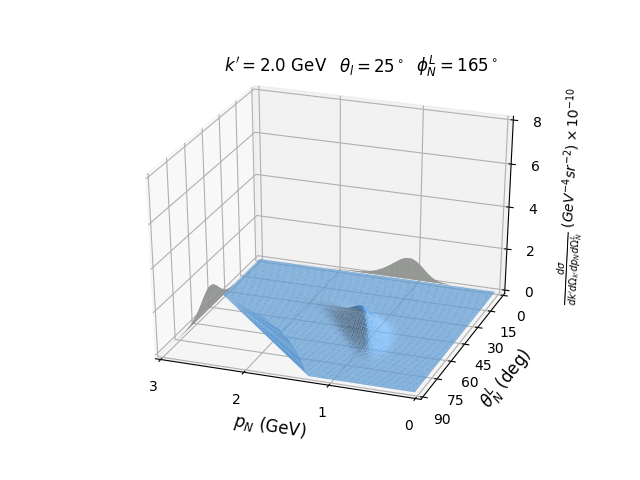}}
	\caption{(Color online) The semi-inclusive cross section for $k'=2\ \mathrm{GeV}$, $\theta_l=25^\circ$ and $\phi_N^L=180^\circ$ (case I)  and $\phi_N^L=165^\circ$ (case II) using the IPSM-RMF spectral function.}\label{fig:semi_RMF}
\end{figure}

Figure \ref{fig:semi_RMF} shows the cross section using the IPSM-RMF spectral function. Its peak is located at $p_N=1.26\ \mathrm{GeV}$ and $\theta_N^L=42.3^\circ$ with a value of $7.58\times 10^{-10}\ \mathrm{GeV}^{-4}\,\mathrm{sr}^{-2}$. The extent of the cross section and its shape a very similar to that for the Rome spectral function for both cases. Note that for case I, the hollow in the peak is also present. This consistent with the assumption that this is associated with the nuclear shell structure. As in the case of the inclusive cross section, multiplying the IPSM-RMF semi-inclusive cross section by 0.88 produces the same peak value as for the Rome result. This, along with limited contributions to the peak of the Rome cross section from large missing momenta and energies suggests that the primary effect on the cross section from the NN interactions included in the Rome spectral function comes from removing strength from lower missing momentum and energy to regions of higher missing momentum and energy. That is, the effect is the same as introducing spectroscopic factors in optical model calculations of the $(e,e'p)$ reaction. The direct effect of short-range correlations will therefore be seen only in regions where the cross sections are much smaller and may be difficult to detect in the CC$\nu$ reaction.

The difference in the inclusive and semi-inclusive cross sections for the Rome spectral function and the IPSM-RMF raises a theoretical issue. The separation energy for CC$\nu$ from $^{16}\mathrm{O}$ is 0.01437 GeV. Use of this separation energy for either of these two models eliminates most, if not all, contributions to the cross sections from the $1p_\frac{1}{2}$ shell. The cross sections shown here have been adjusted to eliminate this problem. 

\begin{figure}
	\centerline{\includegraphics[height=2.5in]{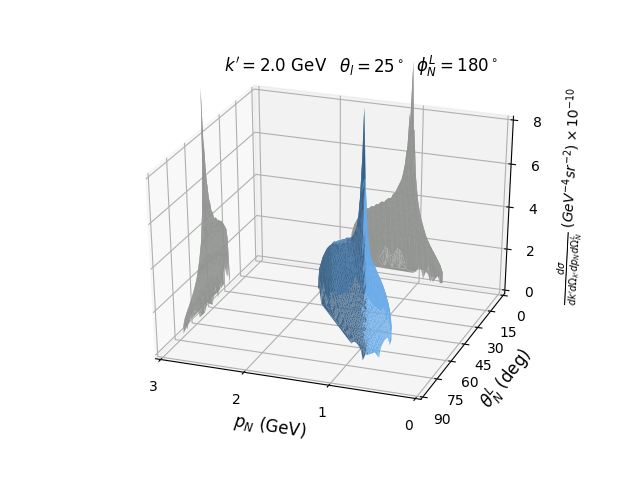}\includegraphics[height=2.5in]{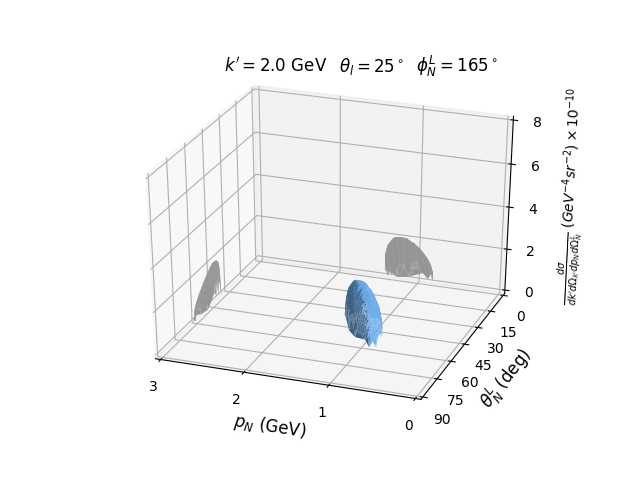}}
	\caption{(Color online) The semi-inclusive cross section for $k'=2\ \mathrm{GeV}$, $\theta_l=25^\circ$ and $\phi_N^L=180^\circ$ (case I) and $\phi_N^L=165^\circ$ (case II) using the LDA spectral function.}\label{fig:semi_LD}
\end{figure}

Figure \ref{fig:semi_LD} shows the cross section using the LDA spectral function. In this case, while the general extent of the cross section is similar to the two previous examples, the shapes of both cases are quite different. Case I contains a spike which is of much greater size than indicated by the figure. This is an artifact of the singularity of the spectral function for missing energy at the separation energy and there is no evidence of the hollow that we have attributed to shell structure for the previous examples. It is not clear how much of this difference would survive the transport mechanisms of the event generators, but it is disturbing.

\begin{figure}
	\centerline{\includegraphics[height=2.5in]{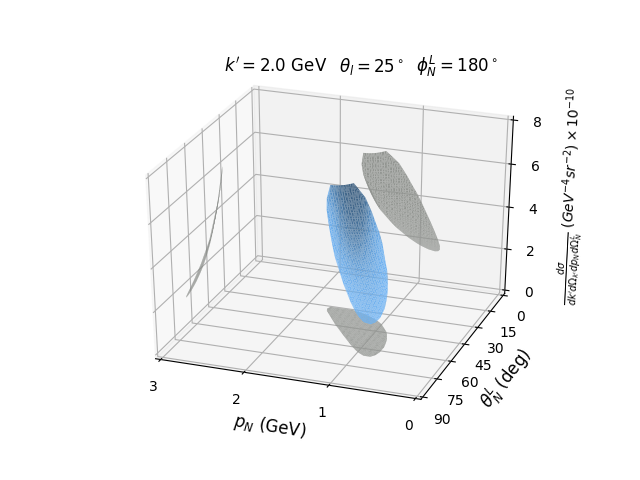}\includegraphics[height=2.5in]{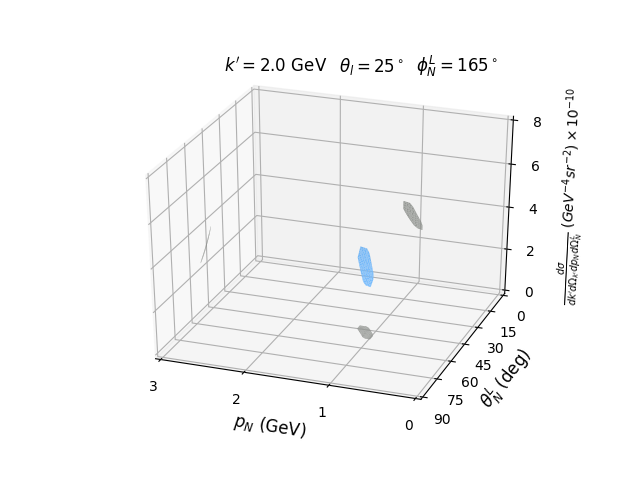}}
	\caption{(Color online) The semi-inclusive cross section for $k'=2\ \mathrm{GeV}$, $\theta_l=25^\circ$ and $\phi_N^L=180^\circ$ (case I) and $\phi_N^L=165^\circ$ (case II) using the RFG spectral function for $k_F=0.23$ GeV.}\label{fig:semi_RFG}
\end{figure}

Finally, Fig.~\ref{fig:semi_RFG} shows two views of the cross section obtained using the RFG spectral function.  Here the maximum value of the cross section for case I is at $p_N=1.41\ \mathrm{GeV}$ and $\theta_N^L=33.3^\circ$ of $5.98\times 10^{-10}\ \mathrm{GeV}^{-4}\,\mathrm{sr}^{-2}$. This is similar in size to the IPSM-RMF and Rome function cross sections. The shape is considerable different from the other models and consists of a simple shell with its extent in the $p_N$-$\theta_N^L$ plane determined by the requirement that $p_m\leq k_F$. An addition projection on to the horizontal plane is provided for clarity.  The considerably different shapes of this model raise questions about the effect that this may have on the output of the event generators.

\begin{figure}
	\centerline{\includegraphics[height=2.5in]{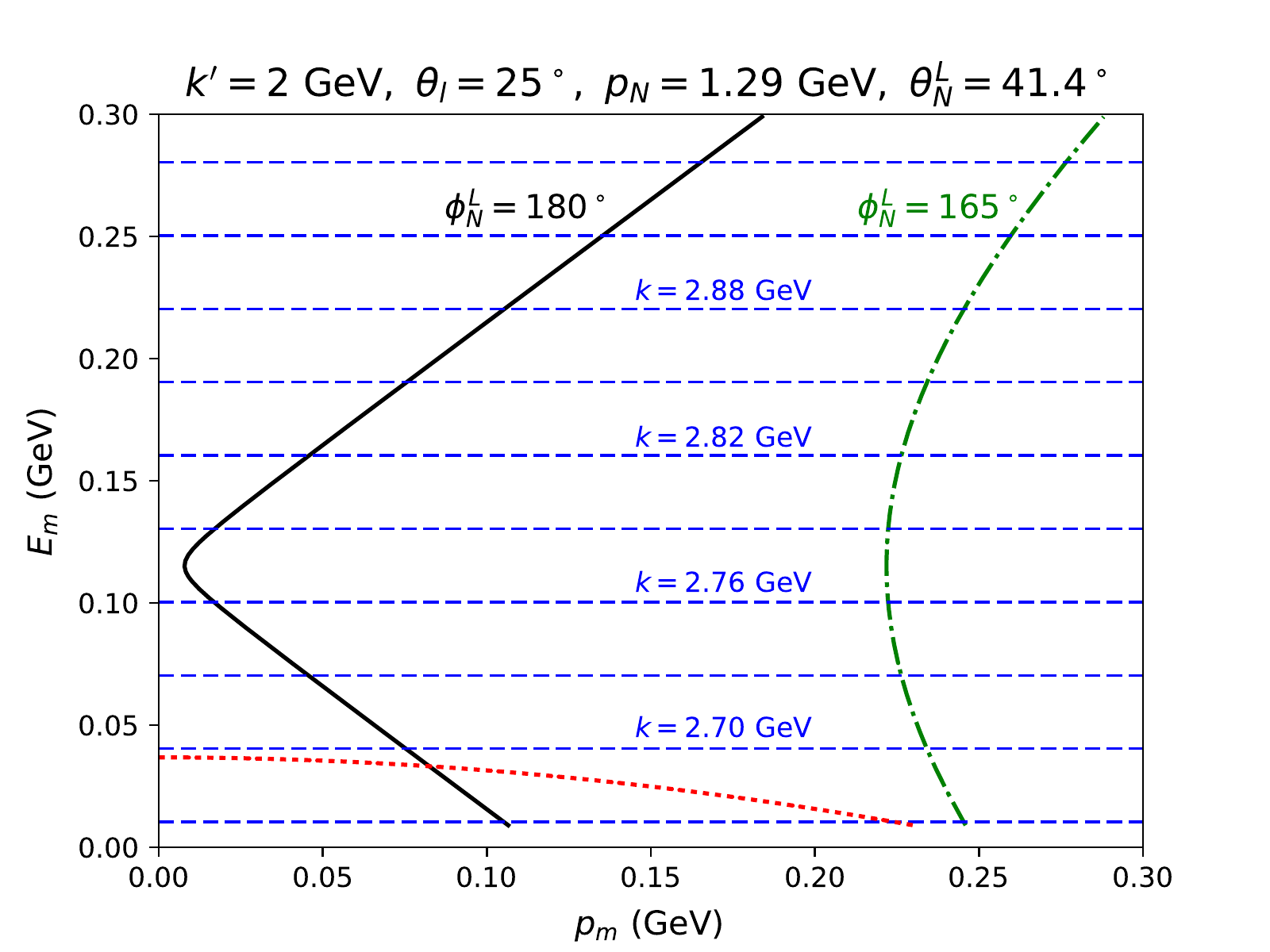}}
	\caption{(Color online) Trajectories for $\phi_N^L=180^\circ$ (solid line) and $\phi_N^L=165^\circ$ (dot-dashed line). The position of the RFG spectral function for $k_F=0.23$ GeV is denoted by the dotted line. Various values of the neutrino energy are indicated by dashed lines.\label{fig:trajectories}}
\end{figure}

The limited extent of the RFG and LDA cross sections is due to very limited regions spanned by the associated spectral functions in the $p_m$-$E_m$ plane. To understand this, note that the neutrino momentum $k_0$ from Eq.~(\ref{eq:k_zero}) is a function of the measured muon and proton energies $\varepsilon'$ and $E_N=\sqrt{p_N^2+m_N^2}$, the known separation energy $E_s$, nucleon mass $m_N$ and the integration variable $\mathcal{E}$. It is independent of the proton angles $\theta_N^L$ and $\phi_N^L$. The missing momentum $p_m$ in Eq.~(\ref{eq:p_missing}) is a function of proton three-momentum $\bm{p}_N^L$, the muon momentum $\bm{k}'$ and $k_0$, which depends implicitly on $\mathcal{E}$. Therefore, the integral over $\mathcal{E}$ in Eq.~(\ref{eq:semi_simple}) evaluates the integrand over a path or trajectory in the $p_m$-$E_m$ plane with $k_0$ varying over each trajectory linearly with $E_m$. Details concerning these trajectories will be addressed a forthcoming paper. Figure \ref{fig:trajectories} shows two trajectories at $\phi_N^L=180^\circ$ and $165^\circ$ for $k'=2$ GeV, $\theta_l=25^\circ$, $p_N=1.29$ GeV and $\theta_N^L=41.4^\circ$. A contribution to the RFG semi-inclusive cross section occurs only when a trajectory crosses the dotted line that indicates the position of the RFG spectral function. This occurs for the $\phi_N^L=180^\circ$ trajectory, but not for the $\phi_N^L=165^\circ$ trajectory. So the RFG will contribute to the semi-inclusive cross section only for the $\phi_N^L=180^\circ$ kinematics. For the LDA, where the largest value of $k_F$ contributing to the spectral function is approximately 0.27 GeV, contributions to the semi-inclusive cross sections will result from integrating along the trajecties up to the position of the RFG spectral function for the larger value of $k_F$, which will extend up to $p_m=0.27$ GeV. For the LDA both of the trajectories in the figure will give finite values. A study of the trajectories associated with points across the hole in case I for the Rome and IPSM-RMF cross sections accounts for this appearance of this feature. 

A useful summary of the semi-inclusive cross sections for the various models is provided in Table~\ref{tab:cross_section_summary}. This gives the position in the $p_N$-$\theta_N^L$ plane for the listed maximum cross section for each of the models employed in this work at both $\phi_N^L=180^\circ$ (case I) and $165^\circ$ (case II) for a muon momentum of 2 GeV and lepton scattering angle $\theta_l=25^\circ$. The row labeled ``Rome*'' refers to the cross section for the Rome spectral function but with all interference response functions set to zero. Note the large size of the LDA cross section, which shows the size of the spike in Fig. \ref{fig:semi_LD}, case I.

\begin{table}
	\begin{tabular}{|l|c|c|l||c|c|r|}\hline\hline
		&\multicolumn{3}{c||}{$\phi_N^L=180^\circ$ (case I)} &\multicolumn{3}{c|}{$\phi_N^L=165^\circ$ (case II)} \\\hline
		& $p_N$ & $\theta_N^L$  & cross section & $p_N$ & $\theta_N^L$ (deg) & cross section \\ \hline
		Rome	&	1.29 GeV &	41.4$^\circ$ &	$6.74\times 10^{-10}$ &	1.11 GeV &	47.7$^\circ$ &	$8.96\times 10^{-11}$ \\
		Rome*	&	1.32 GeV &	42.3$^\circ$ &	$6.42\times 10^{-10}$ &	1.11 GeV &	48.6$^\circ$ &	$8.77\times 10^{-11}$ \\
		IPSM-RMF&	1.26 GeV &	42.3$^\circ$ &	$7.59\times 10^{-10}$ &	1.11 GeV &	47.7$^\circ$ &	$1.03\times 10^{-10}$ \\
		LDA		&	1.11 GeV &	49.5$^\circ$ &	$1.19\times 10^{-5}$ &	1.20 GeV &	45.0$^\circ$ &	$3.78\times 10^{-10}$ \\
		RFG		&	1.41 GeV &	33.3$^\circ$ &	$5.98\times 10^{-10}$ &	1.19 GeV &	45.0$^\circ$ &	$3.75\times 10^{-10}$ \\ \hline\hline		
	\end{tabular}
	\caption{This table contains a summary of $p_N$ and $\theta_N^L$ for the maximum cross sections in $\mathrm{GeV}^{-4}\mathrm{sr}^{-2}$ for the semi-inclusive cross sections shown in Figs. \ref{fig:semi_spectral}--\ref{fig:semi_RFG}. Rome* refers to calculations using the Rome spectral function but excluding all contributions from interference response functions.}\label{tab:cross_section_summary}	
\end{table}

\subsection{Neutrino Energy}\label{sec:Neutrino_Energy}

\begin{figure}
	\centerline{\includegraphics[height=2.5in]{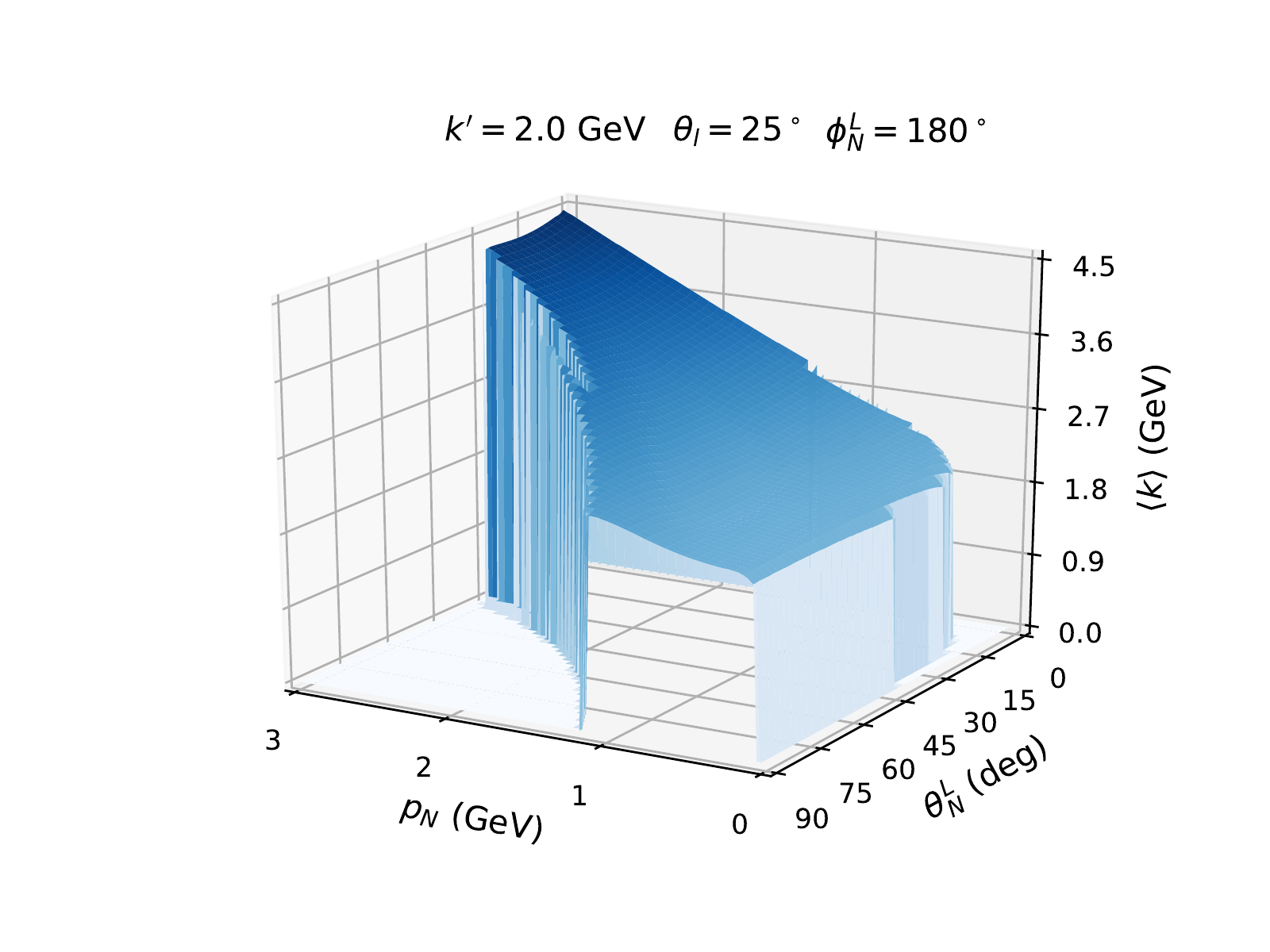}\includegraphics[height=2.5in]{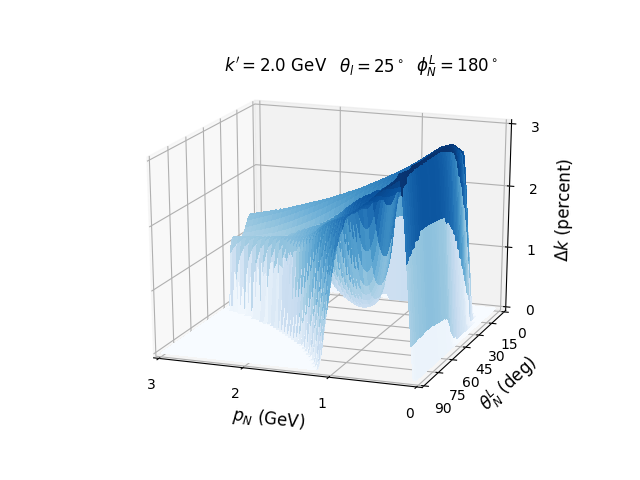}}
	\centerline{\includegraphics[height=2.5in]{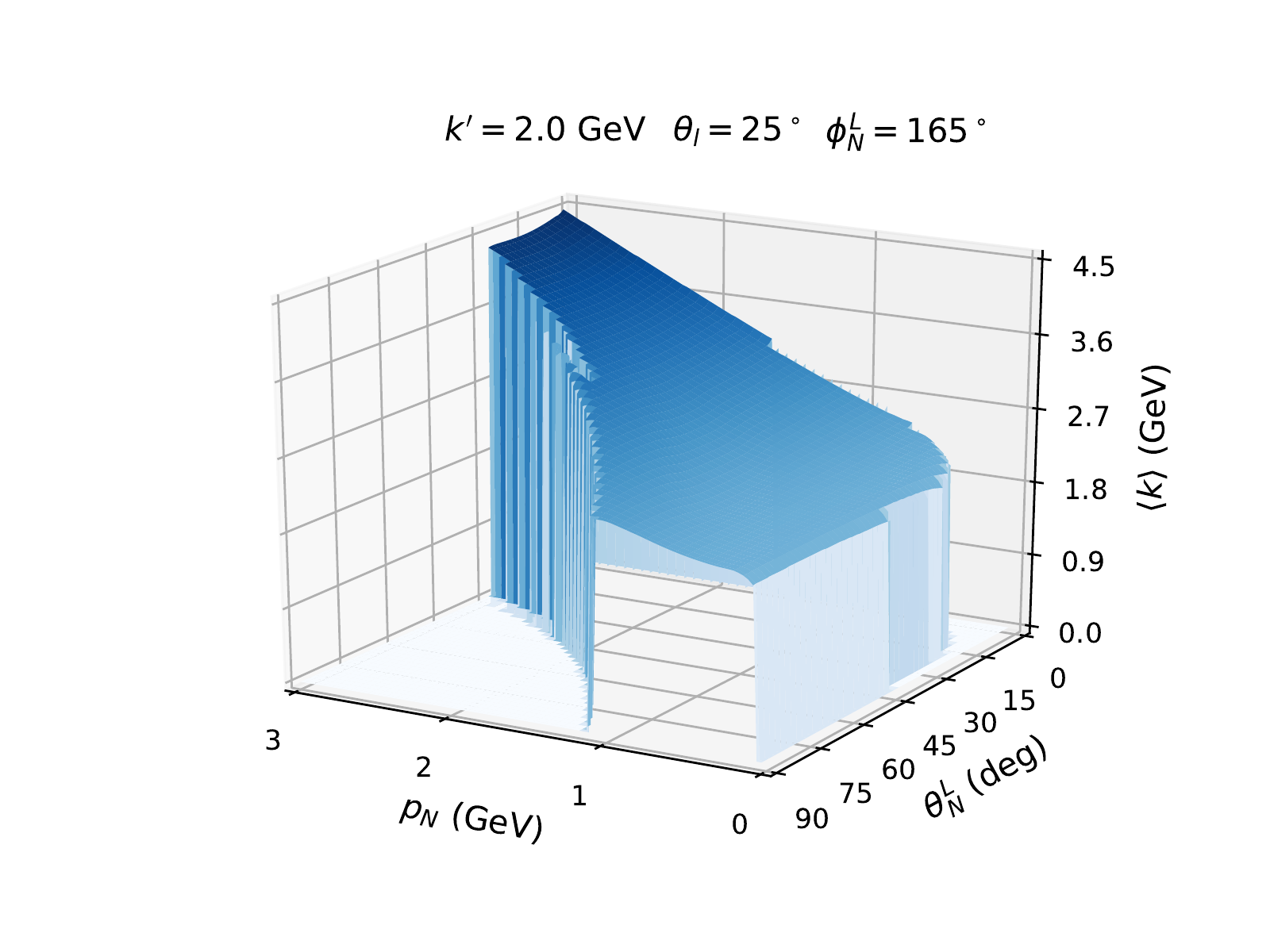}\includegraphics[height=2.5in]{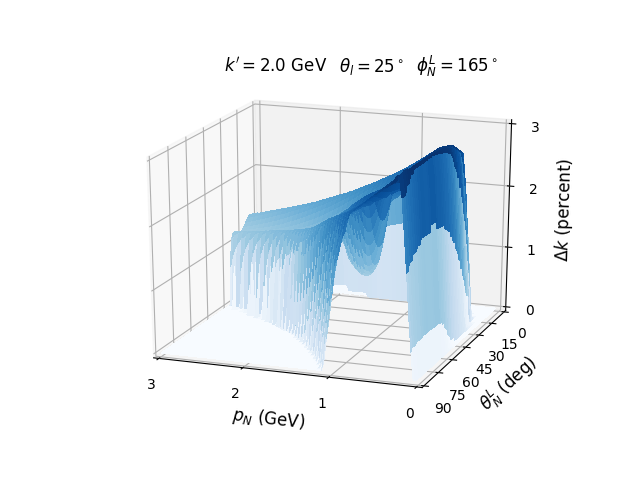}}
	\caption{(Color online) $\left<k\right>$ and $\Delta k$ for the Rome spectral function.}\label{fig:k_Rome}
\end{figure}

Using Eqs.~(\ref{eq:semi_simple})-(\ref{eq:p_missing}) we can define the following moments of the incident neutrino momentum $k$ for each value of the measured kinematic variables as
\begin{align}
\mathcal{K}_n=&\int_0^\infty d\mathcal{E}k_0^{n-1}P(k_0) v_0\widetilde{\mathcal{F}}^{2}_\chi S(p_m,E_s+\mathcal{E})\,.\label{eq:k_moments}
\end{align}
The average momentum at each point is then given by
\begin{equation}
\left<k\right>=\frac{\mathcal{K}_1}{\mathcal{K}_0}\label{eq:k_ave}
\end{equation}
and the standard deviation by
\begin{equation}
\Delta k=\sqrt{\frac{\mathcal{K}_2}{\mathcal{K}_0}-\left<k\right>^2}\,.\label{eq:k_stdev}
\end{equation}
Figure \ref{fig:k_Rome} shows the average momentum and standard deviation calculated using the Rome spectral function under the same conditions as Fig.~\ref{fig:semi_spectral} for $\phi_N^L=180^\circ$ and $165^\circ$. In both cases the $\left<k\right>$ variation is roughly linear with $p_m$ and is only mildly dependent on $\theta_N^L$ within the kinematically allowed region. The right-hand column of plots in this figure show the standard deviation $\Delta k$ for the two azimuthal angles. In both cases this is less than 3\% over the allowed region with the smallest values occurring where the cross section is large and the largest values in the region where the cross section becomes small.

\begin{figure}
	\centerline{\includegraphics[height=2.5in]{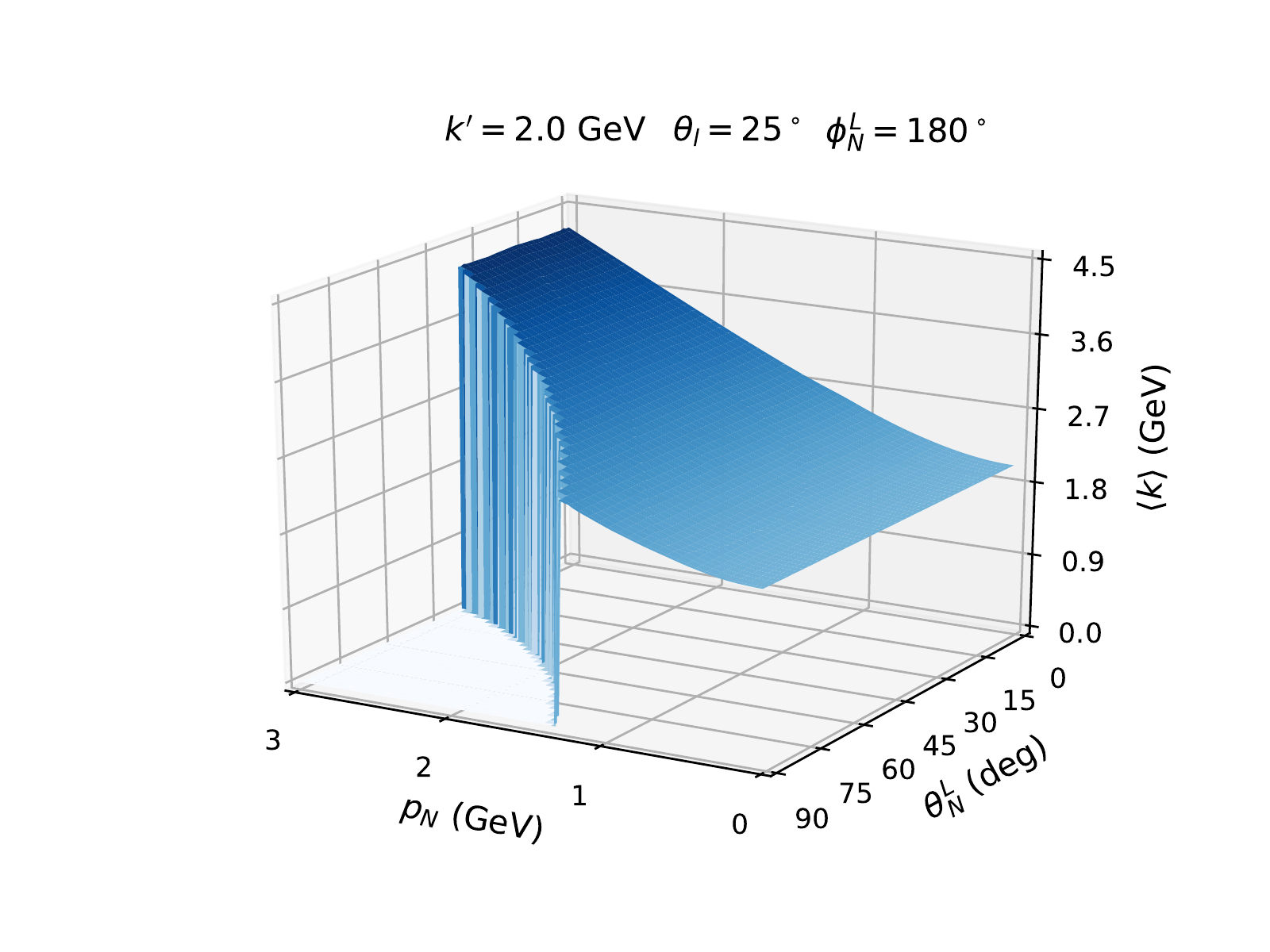}\includegraphics[height=2.5in]{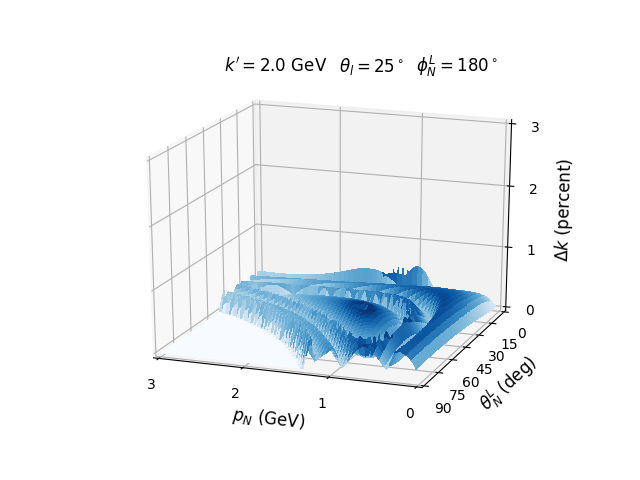}}	\centerline{\includegraphics[height=2.5in]{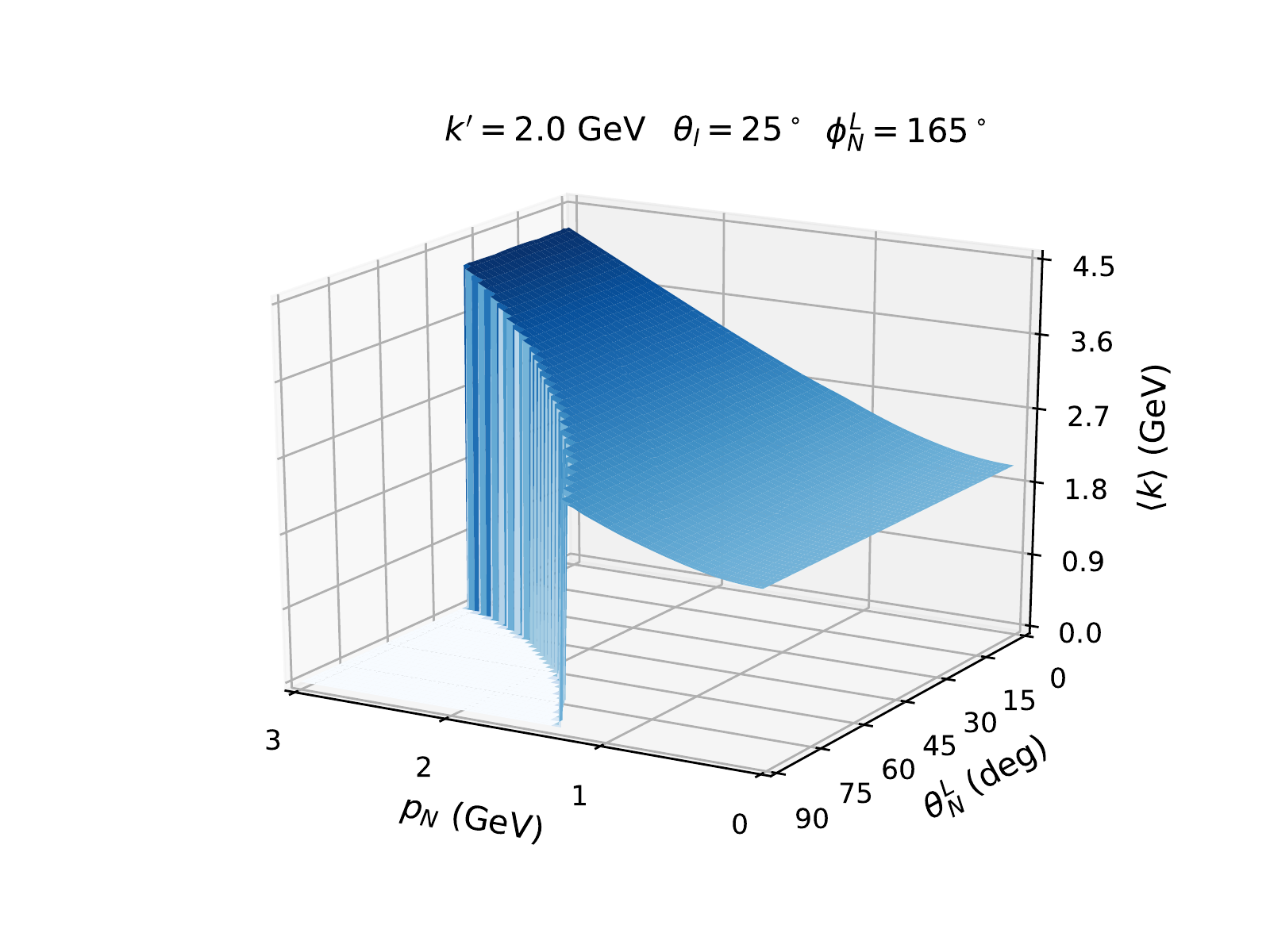}\includegraphics[height=2.5in]{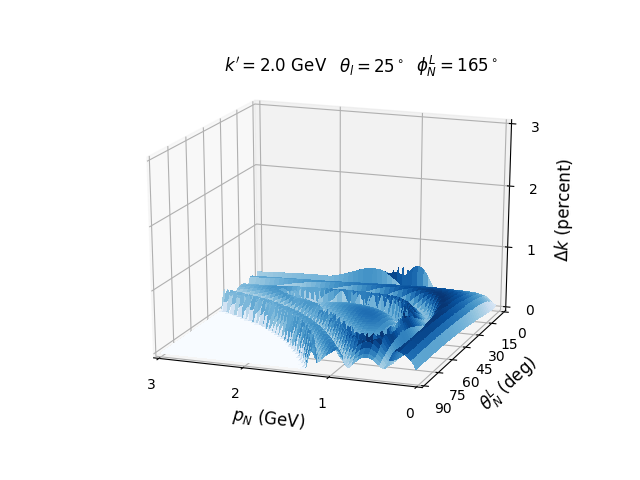}}
	\caption{(Color online) $\left<k\right>$ and $\Delta k$ for the IPSM-RMF spectral function.}\label{fig:k_IPSM}
\end{figure}

Figure \ref{fig:k_IPSM} shows similar plots for the IPSM-RMF model. The average values are similar to those for the Rome spectral function but clearly different in detail. The standard deviations are very small. Here the averages occur only from the three missing energies corresponding to the three delta functions of this spectral function. By averaging only over three values for each kinematic point causes small changes in the average value of $k$ but artificially lowers the values of the standard deviation in comparison with the Rome spectral function where the contributions to the moments of $k$ are distributed over a larger region of $p_m$ and $E_m$.

\begin{figure}
	\centerline{\includegraphics[height=2.5in]{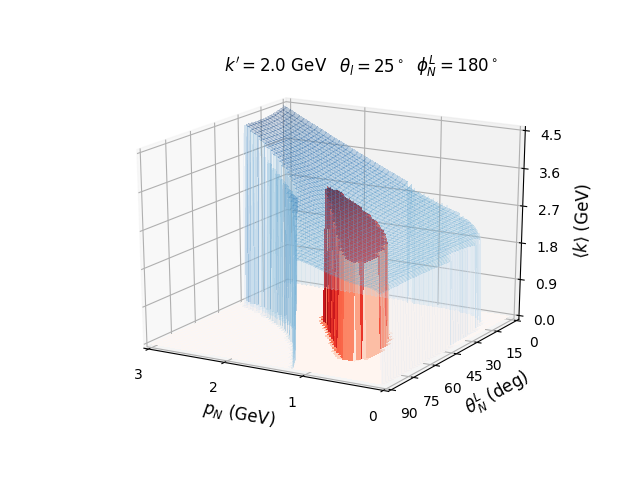}\includegraphics[height=2.5in]{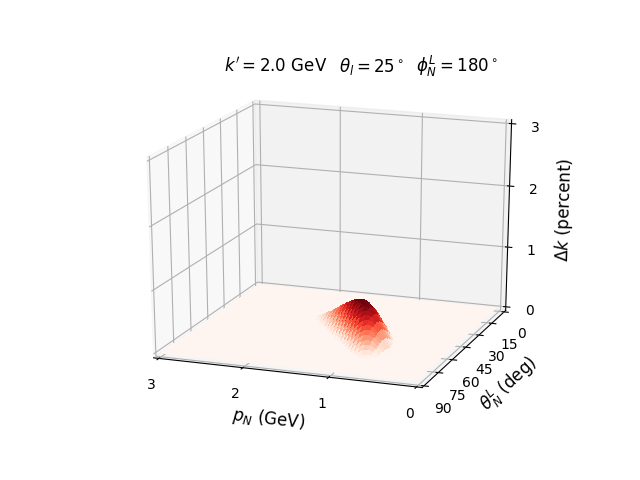}}
	\centerline{\includegraphics[height=2.5in]{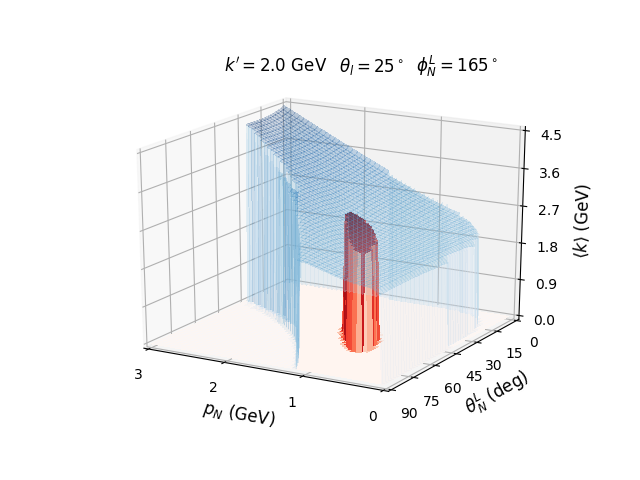}\includegraphics[height=2.5in]{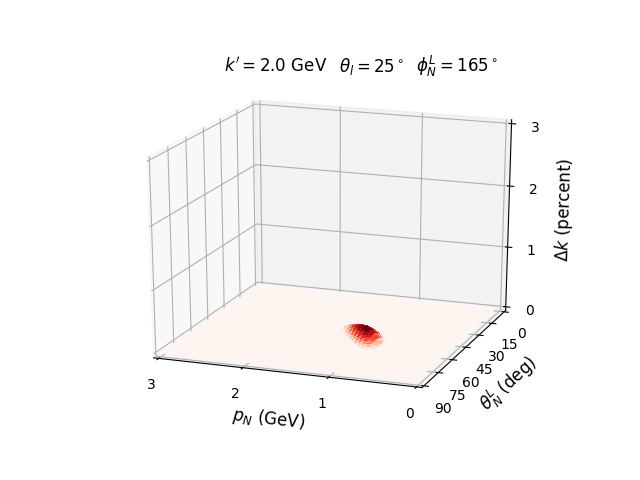}}
	\caption{(Color online) $\left<k\right>$ and $\Delta k$ for the LDA spectral function. Here the plots of $\left< k\right>$ are the darker figures with the the corresponding figures for the Rome model superimposed for purposes of comparison.}\label{fig:k_LDA}
\end{figure}

Figure \ref{fig:k_LDA} shows plots of $\left< k\right>$ and $\Delta k$ for the LDA model.  Here the surfaces for the model are darker with a transparent representation of the results for the Rome model superimposed to allow easy comparison. It should be noted that for $\left< k\right>$, the values for the LDA model are approximately the same as for the Rome model but confined to the limited range in the $p_N$-$\theta_N^L$ plane allowed for this model. The values for $\Delta k$ are considerably different.

\begin{figure}
	\centerline{\includegraphics[height=2.5in]{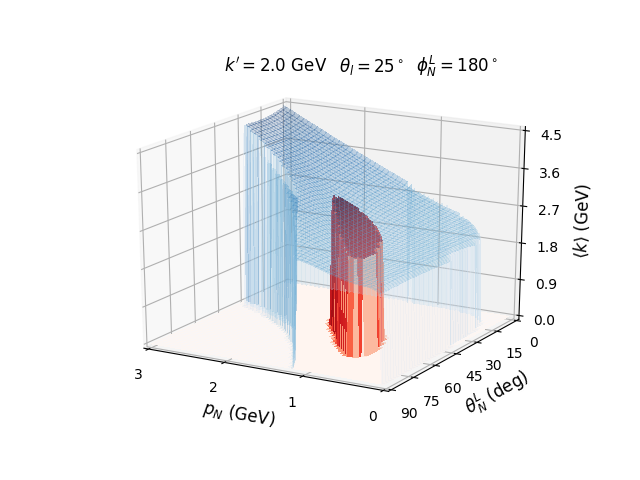}\includegraphics[height=2.5in]{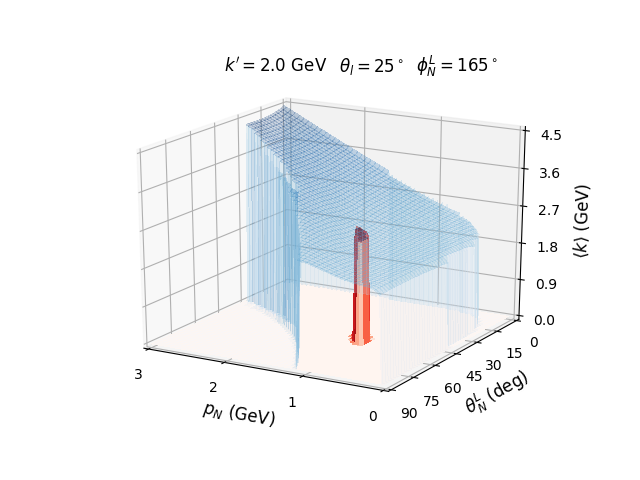}}
	\caption{(Color online) $\left<k\right>$  for the RFG spectral function. The surfaces for $\left< k\right>$ are as described in Fig. \ref{fig:k_LDA}. $\Delta k$ is identically zero for this case.}\label{fig:k_RFG}
\end{figure}

Figure \ref{fig:k_RFG} shows plots of $\left< k\right>$ for the RFG model as plotted in the previous figure.  Again, for $\left< k\right>$, the values for the RFG model are approximately the same as for the Rome model but confined to the limited range in the $p_N$-$\theta_N^L$ plane allowed for this model. $\Delta k$ is identically zero for this model since  the trajectory crosses the RFG spectral function at only one point on the $p_m$-$E_m$ plane for each value in the  $p_N$-$\theta_N^L$ plane.

Figures \ref{fig:k_Rome}-\ref{fig:k_RFG} indicate that values of $k$ for all trajectories for the kinematics shown here vary slowly in comparison with the variation of the spectral functions for each of the models. This suggests that it may be possible to obtain $\left<k\right>$ from semi-inclusive cross sections for models that contain nuclear structure information, such the Rome and IPSM. Further consideration of this use of the semi-inclusive cross sections will be covered in a subsequent paper.

\section{Summary and Conclusions}\label{sec:concl}

In summary we have shown the effect of using four different descriptions of the spectral functions arising from simple independent particle shell model (IPSM) calculation, the relativistic Fermi gas, a local density approximation based on the RFG and the realistic Rome spectral function constructed using a combination of experimental information and many-body theory. While semi-inclusive cross sections the Rome and IPSM calculations are consistent in shape and distribution, they differ in size as would be expected from the application of spectral factors to the shell model calculation, the two calculations based on the RFG differ considerably in shape. In the case of the simple RFG calculation the shape of the semi-inclusive cross section is a simple shell that is significantly different from the other three models. The differences in shape between the Rome and IPSM-RMF semi-inclusive cross sections and the two RFG based models is the result of including the basic shell model features that survive in more sophisticated nuclear many-body calculations.

On the other hand all of this models produce similar results for the inclusive cross section. The lesson from this is that inclusive cross sections provide little indication on the dynamical properties of semi-inclusive cross sections.

The results of Sec. \ref{sec:Neutrino_Energy} suggest that semi-inclusive cross section measurements may provide a reliable method for determining the average value of the incident neutrino energy based on measurable kinematic variables event by event. Further work on this approach will be addressed in a subsequent paper.

The factorable spectral function approach provides a possible method of improving the nuclear physics input into event generators.  It has the advantage that the calculations to produce more accurate spectral functions require only nonrelativistic many-body theory with the relativity isolated to relativistic kinematics for the single-nucleon cross section. It is also possible to use this approach to semi-inclusive cross sections involving the direct production of mesons or single-nucleon DIS cross sections. Techniques for extending this approach to production of two nucleons in the continuum due to two-body currents and short-range correlations are also being developed \cite{Rocco:2018mwt}.

The problems with this approach are associated with the factorization of the cross section. It is necessary to reach a reasonable understanding of where the nuclear physics input should stop and the transport mechanism of the event generators begin. Ultimately, the success of this venture depends on the reliability of the event generator transport mechanisms in reproducing final state interactions which are in sufficient agreement with data.  This issue is to some extent being examined by studies of the accuracy of the event generators in reproducing data obtained from electron scattering in Hall B at Jefferson Lab, where a monoenergetic beam producing larger cross sections with well constrained kinematics can be used to determine the reliability of the event generators \cite{Hall_B}.

\appendix

\section{The Reduced Single-Nucleon Cross Section}\label{sec:appA}

The reduced single-nucleon $CC\nu$ cross section as used in the cross section equations is defined as \cite{semi,VanOrden:2017uyy}

\begin{align}
\widetilde{\mathcal{F}}^2_\chi=& 
\widehat{V}_{CC}
\left(\widetilde{w}^{VV(I)}_{CC}+
\widetilde{w}^{AA(I)}_{CC}\right)+
2\widehat{V}_{CL}
\left(\widetilde{w}^{VV(I)}_{CL}+
\widetilde{w}^{AA(I)}_{CL}\right)+
\widehat{V}_{LL}
\left(\widetilde{w}^{VV(I)}_{LL}
+\widetilde{w}^{AA(I)}_{LL}\right)\nonumber\\
&+ \widehat{V}_{T}
\left(\widetilde{w}^{VV(I)}_{T}
+\widetilde{w}^{AA(I)}_{T}\right)
+\widehat{V}_{TT}
\left(\widetilde{w}^{VV(I)}_{TT}
+\widetilde{w}^{AA(I)}_{TT}\right)\cos 2\phi_N  \nonumber\\
&+ \widehat{V}_{TC}
\left(\widetilde{w}^{VV(I)}_{TC}
+\widetilde{w}^{AA(I)}_{TC}\right)\cos\phi_N+
\widehat{V}_{TL}
\left(\widetilde{w}^{VV(I)}_{TL}+\widetilde{w}^{AA(I)}_{TL}\right)\cos\phi_N \nonumber\\
& 
+\chi \left[ \widehat{V}_{T^{\prime }}
\widetilde{w}^{VA(I)}_{T^{\prime }}
+ \widehat{V}_{TC^{\prime }}
\widetilde{w}^{VA(I)}_{TC^{\prime }}\cos\phi_N
+\widehat{V}_{TL^{\prime }}
\widetilde{w}^{VA(I)}_{TL^{\prime }}\cos\phi_N\right]\,,
\end{align}
where the kinematic coefficients are given by
\begin{align}
\hat{V}_{CC}=&\left(1+\frac{\Delta_1}{v_0}\right)\nonumber\\
\hat{V}_{CL}=&-\frac{1}{|\bm{q}|}\left(\omega+\frac{\Delta_4\kappa}{v_0}\right)\nonumber\\
\hat{V}_{LL}=&\left(\frac{\omega^2}{\bm{q}^2}-\frac{\Delta_1}{v_0}+\frac{\Delta_4^2}{\bm{q}^2v_0}+\frac{2\Delta_4\kappa\omega}{\bm{q}^2v_0}\right)\nonumber\\
\hat{V}_{T}=&\left[|Q^2|\left(\frac{1}{2\bm{q}^2}+\frac{1}{v_0}\right)+\Delta_1\left(\frac{1}{2\bm{q}^2}-\frac{1}{v_0}\right)-\frac{\Delta_1^2-\Delta_3+\Delta_1|Q^2|}{2\bm{q}^2v_0}\right]
\nonumber\\
\hat{V}_{TT}=&-\left[\frac{\Delta_1+|Q^2|}{2\bm{q}^2}\left(1-\frac{\Delta_1}{v_0}\right)+\frac{\Delta_3}{2\bm{q}^2v_0}\right]\nonumber\\
\hat{V}_{TC}=&-\frac{1}{\sqrt{2}v_0}\sqrt{1+\frac{v_0}{\bm{q}^2}}\sqrt{\Delta_3+(\Delta_1+|Q^2|)(v_0-\Delta_1)}\nonumber\\
\hat{V}_{TL}=&\frac{1}{\sqrt{2}\bm{q}^2v_0}\sqrt{\Delta_3+(\Delta_1+|Q^2|)(v_0-\Delta_1)}\left(\Delta_4+\omega\kappa\right)\nonumber\\
\hat{V}_{T'}=&\frac{1}{v_0}\left(|Q^2|\sqrt{1+\frac{v_0}{\bm{q}^2}}-\frac{\Delta_4\omega}{|\bm{q}|}\right)\nonumber\\
\hat{V}_{TC'}=&-\frac{1}{\sqrt{2}v_0}\sqrt{\Delta_3+(\Delta_1+|Q^2|)(v_0-\Delta_1)}\nonumber\\
\hat{V}_{TL'}=&\frac{\omega}{\sqrt{2}|\bm{q}|v_0}\sqrt{\Delta_3+(\Delta_1+|Q^2|)(v_0-\Delta_1)}\nonumber
\end{align}
with
\begin{align}
\Delta_1=&m^2+{m'}^2\nonumber\\
\Delta_2=&2\varepsilon\varepsilon'-2|\bm{k}||\bm{k}'|\nonumber\\
\Delta_3=&4\bm{k}^2\bm{k}^{\prime 2}-4\varepsilon^2\varepsilon^{\prime 2}\nonumber\\
\Delta_4=&m^{\prime 2}-m^2\nonumber\\
\kappa=&\varepsilon+\varepsilon'
\end{align}
and
\begin{align}
v_0=&(\varepsilon+\varepsilon')^2-\bm{q}^2\nonumber\\
=&\Delta_2
+\Delta_1+4|\bm{k}||\bm{k}'|\cos^2\frac{\theta_l}{2}\,.
\nonumber
\end{align}

The angle $\phi_N$ is the azimuthal angle of $\bm{p}_N$ about the three-momentum transfer $\bm{q}$. This can be obtained from the lab frame kinematical variables by defining the angle between the neutrino beam direction and $\bm{q}$ as
\begin{equation}
\theta_q=\cos^{-1}\left(\frac{k-k'\cos\theta_l}{q}\right)\,.
\end{equation}
The polar angle of $\bm{p}_N$ is given by
\begin{equation}
\theta_N=\cos^{-1}(\cos\theta_N^L\cos\theta_q-\cos\phi_N^L\sin\theta_N^L\cos\theta_q)\,.
\end{equation}
The cosine and sine of $\phi_N$ are then
\begin{equation}
\cos\phi_N=\frac{\cos\phi_N^L\sin\theta_N^L\cos\theta_q+\cos\theta_N^2\sin\theta_q}{\sin\theta_N}
\end{equation}
and
\begin{equation}
\sin\phi_N=\frac{\sin\phi_N^L\sin\theta_N^L}{\sin\theta_N}\,.
\end{equation}
\begin{align}
8 m_N^4\widetilde{w}^{VV(I)}_{CC}=&4 E_p^2 \left(4 F_1^2(|Q|^2)
m_N^2+F^2_2(|Q|^2)
|Q|^2\right)+4 E_p \omega 
\left(4 F_1^2(|Q|^2)
m_N^2+F^2_2(|Q|^2)
|Q|^2\right)\nonumber\\
&-4 F_1^2(|Q|^2)
m_N^2 |Q|^2-8 F_1(|Q|^2)
F_2(|Q|^2) m_N^2 \left(\omega
^2+|Q|^2\right)\nonumber\\
&+F^2_2(|Q|^2)
\left(\omega ^2 |Q|^2-4 m_N^2
\left(\omega ^2+|Q|^2\right)\right)\nonumber\\
&-2\delta (2 E_p+\omega ) \left(F^2_2(|Q|^2)
\left(2 E_p \omega +\omega
^2-|Q|^2\right)-4 F_1^2(|Q|^2)
m_N^2\right)\nonumber\\
&+\delta^2\left(-4  E_p^2 F^2_2(|Q|^2)-12 E_p
F^2_2(|Q|^2) \omega +4 F_1^2(|Q|^2)
m_N^2+F^2_2(|Q|^2)\right.\nonumber\\
&\left.\times\left(|Q|^2-5 \omega ^2\right)\right)-4\delta^3 F^2_2(|Q|^2) (E_p+\omega )-\delta^4F^2_2(|Q|^2)
\end{align}
\begin{align}   
8 m_N^4\widetilde{w}^{AA(I)}_{CC}=&16 E_p^2 G^2_A(|Q|^2)
m_N^2+16 E_p G^2_A(|Q|^2)
m_N^2 \omega -4 G^2_A(|Q|^2)
m_N^2 \left(4
m_N^2+|Q|^2\right)\nonumber\\
&-8
G_A(|Q|^2) G_P(|Q|^2) m_N^2 \omega
^2+G^2_P(|Q|^2) \omega ^2 |Q|^2 \nonumber\\
&+\delta\left(16 E_p G^2_A(|Q|^2) m_N^2+8
G^2_A(|Q|^2) m_N^2 \omega\right.\nonumber\\
&\left. -8
G_A(|Q|^2) G_P(|Q|^2) m_N^2 \omega
-2 G^2_P(|Q|^2) \omega ^3\right)\nonumber\\
& +\delta^2\left(4 G^2_A(|Q|^2) m_N^2-G^2_P(|Q|^2)
\omega ^2\right)
\end{align}
\begin{align}   
8 m_N^4\widetilde{w}^{VV(I)}_{CL}=&2 E_p (2 p_\parallel+q) \left(4
F_1^2(|Q|^2) m_N^2+F^2_2(|Q|^2)
|Q|^2\right)+\omega  \left(8
F_1^2(|Q|^2) m_N^2 p_\parallel\right.\nonumber\\
&\left. -8
F_1(|Q|^2) F_2(|Q|^2) m_N^2
q+F^2_2(|Q|^2) \left(-4 m_N^2
q+2 p_\parallel |Q|^2+q
|Q|^2\right)\right)\nonumber\\
&\delta\left[\left(F^2_2(|Q|^2) \left(-\left(4 E_p^2
q+E_p \omega  (4 p_\parallel+6
q)+2 \omega ^2 (p_\parallel+q)-|Q|^2
(2 p_\parallel+q)\right)\right)\right.\right.\nonumber\\
&\left.\left.+8
F_1^2(|Q|^2) m_N^2 p_\parallel-4
F_1(|Q|^2) F_2(|Q|^2) m_N^2 q\right)\right]\nonumber\\
&-\delta^2F^2_2(|Q|^2) (4 E_p q+2 \omega 
p_\parallel+3 \omega  q)\nonumber\\
&-\delta^3F^2_2(|Q|^2) q
\end{align}
\begin{align}   
8 m_N^4\widetilde{w}^{AA(I)}_{CL}=&8 E_p G^2_A(|Q|^2) m_N^2 (2
p_\parallel+q)+\omega  \left(8
G^2_A(|Q|^2) m_N^2 p_\parallel-8
G_A(|Q|^2) G_P(|Q|^2) m_N^2
q\right.\nonumber\\
&\left.+G^2_P(|Q|^2) q |Q|^2\right)\nonumber\\
&\delta\left(8 G^2_A(|Q|^2) m_N^2 p_\parallel-4
G_A(|Q|^2) G_P(|Q|^2) m_N^2 q-2
G^2_P(|Q|^2) \omega ^2 q\right)\nonumber\\
&-\delta^2G^2_P(|Q|^2) \omega  q
\end{align}
\begin{align}   
8 m_N^4\widetilde{w}^{VV(I)}_{LL}=&16 F_1^2(|Q|^2) m_N^2 p_\parallel
(p_\parallel+q)-8 F_1(|Q|^2) F_2(|Q|^2)
m_N^2 q^2\nonumber\\
&+F^2_2(|Q|^2)
\left(|Q|^2 (2 p_\parallel+q)^2-4
m_N^2 q^2\right)-2\delta F^2_2(|Q|^2) q (2 E_p+\omega ) (2
p_\parallel+q)\nonumber\\
&-\delta^2 F^2_2(|Q|^2) q (4 p_\parallel+q)
\end{align}
\begin{align}   
8 m_N^4\widetilde{w}^{AA(I)}_{LL}=&16 G^2_A(|Q|^2) m_N^2 p_\parallel
(p_\parallel+q)-8 G_A(|Q|^2) G_P(|Q|^2)
m_N^2 q^2+G^2_P(|Q|^2) q^2
|Q|^2\nonumber\\
&-2\delta G^2_P(|Q|^2) \omega  q^2-\delta^2G^2_P(|Q|^2) q^2
\end{align}
\begin{align}   
8 m_N^4\widetilde{w}^{VV(I)}_{T}=&4 (4 F_1(|Q|^2) F_2(|Q|^2) m_N^2 |Q^2| + F_2^2(|Q|^2) (2 m_N^2 + p_\perp^2) |Q^2| \nonumber\\
&+ 
2 F_1^2(|Q|^2) m_N^2 (2 p_\perp^2 + |Q^2|))-16\delta F_1(|Q|^2) m_N^2 \omega\nonumber\\
& +\delta^2 \left[8 E_p^2 F^2_2(|Q|^2)+8 E_p
F^2_2(|Q|^2) \omega -8 F_1^2(|Q|^2)
m_N^2\right.\nonumber\\
&\left.-2 F^2_2(|Q|^2) |Q|^2
(F_1(|Q|^2)+F_2(|Q|^2))\right]+4\delta^3 F^2_2(|Q|^2) (2 E_p+\omega ) \nonumber\\
&+2\delta^4 F^2_2(|Q|^2)
\end{align}
\begin{align}   
8 m_N^4\widetilde{w}^{AA(I)}_{T}=&8 G^2_A(|Q|^2) m_N^2 \left(4
m_N^2+2
p_\perp^2+|Q|^2\right)-16\delta G^2_A(|Q|^2) m_N^2 \omega\nonumber\\
&-8\delta^2 G^2_A(|Q|^2) m_N^2
\end{align}
\begin{align}   
8 m_N^4\widetilde{w}^{VV(I)}_{TT}=&-4 p_\perp^2 \left(4 F_1^2(|Q|^2)
m_N^2+F^2_2(|Q|^2)
|Q|^2\right)\\
8 m_N^4\widetilde{w}^{AA(I)}_{TT}=&-16 G^2_A(|Q|^2) m_N^2
p_\perp^2
\end{align}
\begin{align}   
8 m_N^4\widetilde{w}^{VV(I)}_{TC}=&4 \sqrt{2} p_\perp (2 E_p+\omega
) \left(4 F_1^2(|Q|^2)
m_N^2+F^2_2(|Q|^2)
|Q|^2\right)\nonumber\\
&+4 \sqrt{2}\delta p_\perp
\left(F^2_2(|Q|^2) \left(-2 E_p
\omega -\omega ^2+|Q|^2\right)+4
F_1^2(|Q|^2) m_N^2\right)\nonumber\\
&-4\delta^2 \sqrt{2} F^2_2(|Q|^2) \omega 
p_\perp
\end{align}
\begin{align}   
8 m_N^4\widetilde{w}^{AA(I)}_{TC}=&16 \sqrt{2} G^2_A(|Q|^2) m_N^2
p_\perp (2 E_p+\omega )+16 \sqrt{2}\delta G^2_A(|Q|^2) m_N^2
p_\perp
\end{align}
\begin{align}   
8 m_N^4\widetilde{w}^{VV(I)}_{TL}=&4 \sqrt{2} p_\perp (2
p_\parallel+q) \left(4 F_1^2(|Q|^2)
m_N^2+F^2_2(|Q|^2)
|Q|^2\right)\nonumber\\
&-4 \sqrt{2}\delta F^2_2(|Q|^2) p_\perp q
(2 E_p+\omega )-4 \sqrt{2}\delta^2 F^2_2(|Q|^2) p_\perp q
\end{align}
\begin{align}   
8 m_N^4\widetilde{w}^{AA(I)}_{TL}=&16 \sqrt{2} G^2_A(|Q|^2) m_N^2
p_\perp (2 p_\parallel+q)
\end{align}
\begin{align}   
8 m_N^4\widetilde{w}^{VA(I)}_{T'}=&-32 G_A(|Q|^2) m_N^2
(F_1(|Q|^2)+F_2(|Q|^2)) (\omega 
p_\parallel-E_p q)\nonumber\\
&-16\delta G_A(|Q|^2) m_N^2 (2 F_1(|Q|^2)
p_\parallel-F_2(|Q|^2) q)
\end{align}
\begin{align}   
8 m_N^4\widetilde{w}^{VA(I)}_{TC'}=&-32 \sqrt{2} G_A(|Q|^2) m_N^2
p_\perp q (F_1(|Q|^2)+F_2(|Q|^2))\nonumber\\
&-4 \sqrt{2}\delta F_2(|Q|^2) G_P(|Q|^2) \omega 
p_\perp q
\end{align}
\begin{align}   
8 m_N^4\widetilde{w}^{VA(I)}_{TL'}=&32 \sqrt{2} G_A(|Q|^2) m_N^2 \omega 
p_\perp (F_1(|Q|^2)+F_2(|Q|^2))\nonumber\\
&+4 \sqrt{2}\delta p_\perp \left(8
F_1(|Q|^2) G_A(|Q|^2)
m_N^2-F_2(|Q|^2) G_P(|Q|^2)
q^2\right)\,,
\end{align}
where
\begin{align}
p_\parallel&=\frac{\bm{p}\cdot\bm{q}}{q}\\
p_\perp&=\frac{|\bm{p}\times\bm{q}|}{q}\\
E_p&=\sqrt{p^2+m_N^2}\\
|Q^2|&=q^2-\omega^2\,,
\end{align}
and
\begin{equation}
	\delta=\sqrt{p_N^2+m_N^2}-\sqrt{p_m^2+m_N^2}-\omega\,.
\end{equation}
The isovector electromagnetic form factors $F_1$ and $F_2$ are from \cite{FF_GK_05_2_Lomon_02,FF_GK05_1_Lomon_06} and the weak form factors $G_A$ and $G_P$ are simple dipole forms as used in \cite{deut}.

\begin{acknowledgments}
This work has been supported in part by the Office of Nuclear Physics of the US Department of Energy under Grant Contract DE-FG02-94ER40818 (T. W. D.), by the US Department of Energy under Contract No. DE-AC05-06OR23177, and by Jefferson Science Associates, LLC under U.S. DOE Contract DEAC05-06OR23177 and by U.S. DOE Grant DE-FG02-97ER41028 (J. W. V. O.). The authors also thank Omar Benhar for kindly providing a model for the spectral function of ${}^{16}$O employed in this study.
\end{acknowledgments}

\bibliography{neutrino}

\end{document}